\documentclass[aps,prd,amssymb,amsmath,amsfonts,superscriptaddress,nofootinbib,eqsecnum,reprint,showpacs,longbibliography]{revtex4-2}

\usepackage{graphicx}
\usepackage{lmodern}
\usepackage{amsmath,amssymb}
\usepackage{mathrsfs}
\usepackage{amsfonts}
\usepackage[utf8]{inputenc}
\usepackage{url}
\usepackage[colorlinks]{hyperref}
\usepackage{xcolor}
\usepackage[normalem]{ulem}
\usepackage{orcidlink}
\usepackage{placeins}
\usepackage{mathtools}
\usepackage{rotating}
\usepackage{tabularx}
\usepackage{enumitem} 
\usepackage{xspace}
\usepackage{subcaption}
\usepackage{multirow}
\usepackage{booktabs}
\usepackage[justification=justified,singlelinecheck=false]{caption}

\newcommand{\sat}{\mathrm{sat}}
\newcommand{\sym}{\mathrm{sym}}

\definecolor{dodgerblue}{HTML}{1E90FF}
\definecolor{NatalieGreen}{HTML}{336600}
\definecolor{AnnaOrange}{HTML}{e67300}
\definecolor{GuilieBlue}{HTML}{000066}
\definecolor{TimPurple}{HTML}{8303a6}
\definecolor{cesarRed}{HTML}{FF0000}

\hypersetup{
     colorlinks=true,
     linkcolor=dodgerblue,
     filecolor=dodgerblue,
     citecolor = dodgerblue,      
     urlcolor=dodgerblue,
     }

\newcommand{\Potsdam}{Universit\"at Potsdam, Institut f\"ur Physik und Astronomie, \\
Haus 28, Karl-Liebknecht-Str. 24/25, 14476, Potsdam, Germany}

\begin{abstract}
    Gravitational-wave observations of binary neutron star systems can shed light on the currently unknown dense matter equation of state. The equation of state determines a large number of neutron star properties, such as tidal deformability, radius, and quadrupole moment, several of which directly affect the emitted gravitational-wave signals. To reduce the dimensionality when computing gravitational-waves and when interpreting observational data, quasi-universal relations are commonly employed to connect different neutron star properties. However, quasi-universal relations are not exact and their use may introduce uncertainty and bias. We explore the potential biases arising from different quasi-universal relations in the third generation era: (i) the Love-Q relation connecting the spin-induced quadrupole moment and the tidal deformability, (ii) the relation between the fundamental mode frequency and the tidal deformability, and (iii) the binary Love relation. We find that for the quadrupole relation biases are only present for rapidly rotating systems, for the binary-Love relation induces moderate biases only in the next-to-leading-order tidal parameters, which can however propagate into the inferred equation of state at low masses. Regarding fundamental mode frequencies, we find that the employed relation introduces only negligible biases, while waveform systematic effects can become comparatively large. Our results highlight that while quasi-universal relations remain a useful tool within gravitational-wave analyses, careful treatment is needed to avoid biases in equation of state measurements with next-generation detectors.
\end{abstract}

\DeclarePairedDelimiterX{\norm}[1]{\lVert}{\rVert}{#1}

\begin{document}

\title{Investigating the impact of quasi-universal relations on neutron star constraints in third-generation detectors}

\author{Natalie Williams \orcidlink{0000-0002-5656-8119}} 
\affiliation{\Potsdam}

\author{Anna Puecher  \orcidlink{0000-0003-1357-4348}} 
\affiliation{\Potsdam}

\author{Guilherme Grams 
\orcidlink{0000-0002-8635-383X}}
\affiliation{\Potsdam}

\author{César V. Flores
\orcidlink{0000-0003-1298-9920}}
\affiliation{Universidade Estadual da Região Tocantina do Maranhão, UEMASUL, Centro de Ciências Exatas, Naturais e Tecnológicas, Imperatriz, CEP 65901-480 Maranhão, Brazil}
\affiliation{
Universidade Federal do Maranhão, Programa de Pós-graduação em Física, 65080-805, São Luís, Maranhão, Brazil}

\author{Tim Dietrich \orcidlink{0000-0003-2374-307X}} 
\affiliation{\Potsdam}
\affiliation{Max Planck Institute for Gravitational Physics (Albert Einstein Institute), \\ Am M\"uhlenberg 1, Potsdam 14476, Germany}

\date{\today}

\maketitle

\section{Introduction}
\label{sec:intro}
Understanding the interior composition of neutron stars (NS) remains a key challenge in nuclear astrophysics. Gravitational waves (GWs) from binary neutron star (BNS) mergers provide a means of accessing the ultra-dense regime in NS cores and shedding light on the still unknown dense matter equation of state (EOS). Tidal interactions between the two NS modify the orbital dynamics, leaving a characteristic imprint on the gravitational waveform, and in turn encoding information about the EOS. This tidal signature was first observed in GW170817\cite{LIGOScientific:2017vwq, LIGOScientific:2017zic, LIGOScientific:2017ync} as detected with Advanced LIGO~\cite{LIGOScientific:2014pky} and Advanced Virgo~\cite{VIRGO:2014yos}, providing the first GW constraints on the dense-matter EOS~\cite{LIGOScientific:2018cki, LIGOScientific:2018hze, De:2018uhw, Raithel:2018ncd}.

The effect of matter on the GW waveform is primarily characterised by the quadrupole \textit{tidal deformability} $\Lambda=(2/3)k_2(R/m)^5$, which depends on the tidal Love number $k_2$, mass $m$, and radius $R$ of the NS. The tidal deformability describes the deformation of the NS due to the tidal field induced by its companion~\cite{Flanagan:2007ix, Hinderer:2007mb}. However, unlike black holes, which are fully described by mass, spin, and charge under the no-hair theorem, NSs possess a rich array of EOS-dependent physical properties beyond the tidal deformability that create additional imprints on the GW signal. These include a broad spectrum of oscillation modes~\cite{Kokkotas:1999bd}, non-constant spin induced quadrupole moments~\cite{Hartle:1968si, Laarakkers:1997hb}, higher order tides beyond the quadrupole~\cite{Akcay:2018yyh} and current multipolar effects~\cite{Damour:2009vw}.

Incorporating these physical effects is necessary for an accurate and complete description of the waveform. Accurate modeling ensures that source properties, and consequenctly EOS constraints, can be reliably extracted from the data. However, including matter effects beyond tidal deformability would increase the dimensionality of the parameter space. This poses a challenge for both waveform modeling and parameter estimation, particularly given the high computational cost in the analysis of long-duration BNS signals~\cite{Hu:2024mvn}.

These matter effects all depend sensitively on the EOS. However, it has been shown that a set of \textit{quasi-universal relations} (qURs) exist, e.g., Ref.~\cite{Yagi:2013bca}, connecting sets of these properties to one another. Several underlying mechanisms for this approximate universality have been proposed~\cite{Yagi:2014qua, Sham:2014kea, Kyutoku:2025zud}, with recent work suggesting that universality is attributed to the low compressibility of NS matter~\cite{Katagiri:2025qze}. Critically, qURs are not only conceptually intriguing, they are also a useful tool for modeling NSs. Given a mass and a tidal deformability, qURs enable estimation of a wide range of NS properties without expanding the parameter space. Consequently, qURs are often employed within waveform models and parameter estimation to improve performance.

Prominently among qURs are the \textit{I-Love-Q} relations introduced by Yagi and Yunes~\cite{Yagi:2013bca, Yagi:2013awa}, which link the moment of inertia, tidal deformability, and spin-induced quadrupole moment for slowly rotating isolated NSs in general relativity. Subsequent work has since extended these relations, e.g. to higher multipole moments~\cite{Yagi:2013sva}, rapid rotation~\cite{Pappas:2013naa, Chakrabarti:2013tca, Kruger:2023olj, Doneva:2014faa}, strong magnetic fields~\cite{Haskell:2013vha}, exotic matter~\cite{Kumar:2023ojk, Roy:2025xdw, Pani:2015tga, Uchikata:2016qku} and beyond general relativity~\cite{Yagi:2013awa, Yagi:2013bca, Yunes:2009ch,Yagi:2012ya,Yagi:2013mbt,Kleihaus:2016dui,Kleihaus:2014lba,Doneva:2014faa,Pani:2014jra,Doneva:2016xmf,Doneva:2015hsa,Staykov:2015cfa,Sham:2013cya,Pani:2015tga,Uchikata:2016qku}. Closely related is the Love–C relation~\cite{Maselli:2013mva,Jiang:2020uvb, Lowrey:2024anh}, which connects the tidal deformability to the stellar compactness and enables constraints on NS radii from tidal measurements. Analogous qURs relate oscillation frequencies - such as those corresponding to the fundamental (\textit{f}) and gravity (\textit{g})  modes - to the tidal deformability~\cite{Lau:2009bu, Chan:2014kua, Chirenti:2015dda, Sotani:2021kiw, Zhao:2022tcw}. Additionally, in the binary context, the binary-Love qURs~\cite{Yagi:2016qmr} connect the tidal deformabilities of the two stars in a binary system. Applications of qURs include obtaining constraints on NS quantities~\cite{Kumar:2019xgp, Godzieba:2020bbz, Chatterjee:2025pkx, Ghosh:2023vja}, breaking degeneracies in GW parameter estimation~\cite{Xie:2022brn, Chatziioannou:2018vzf}, measuring the Hubble constant~\cite{Chatterjee:2021xrm}, and performing direct measurements of the relations themselves~\cite{Samajdar:2020xrd, Zheng:2025nij}.

However, qURs remain inherently approximate, and the degree of EOS dependence varies across qURs. For example, the qUR between the tidal deformability and the moment of inertia carries error up to $\mathcal{O}(1\%)$~\cite{Yagi:2013bca}, however, the binary Love qUR exhibits errors of up to $\mathcal{O}(10\%)$~\cite{Yagi:2015pkc}. Strong first order phase transitions in the EOS, signaling the presence of exotic matter, are known to produce deviations from qURs~\cite{Bauswein:2018bma}. More generally, it has been shown that deviations from qURs arise for EOSs exhibiting large mass-radius variations, including hadronic EOSs with extreme slope in the mass-radius curve~\cite{Raithel:2022orm}. 

Given the sub-dominant nature of matter effects beyond the tidal deformability for typical low-spin BNS systems, it is unexpected that any systematic errors from qUR deviations would exceed statistical errors given current detector sensitives~\cite{Godzieba:2020bbz}. However, third generation detectors such as the Einstein Telescope (ET)~\cite{Punturo:2010zz, Branchesi:2023mws, ET:2019dnz, ET:2025xjr} and Cosmic Explorer~\cite{Evans:2021gyd, Reitze:2019iox} forecast much improved sensitivity, predicting an era of $\sim10^4$ BNS detections per year~\cite{Baibhav:2019gxm} and exquisite EOS constraints~\cite{Pacilio:2021jmq,Iacovelli:2023nbv,Walker:2024loo,Gupta:2022qgg,Ghosh:2022muc,Finstad:2022oni,Pradhan:2023zor,Pradhan:2023xtq}. Such a leap in precision makes it imperative to carefully examine the use of qURs within analyses, or risk biasing future EOS measurements~\cite{Kashyap:2022wzr, Suleiman:2024ztn}.

In this paper, we assess a subset of qURs for two EOSs with extreme mass-radius slopes to determine whether they remain reliable for use within GW analyses with the sensitivity of third-generation detectors. We investigate binary systems with parameters chosen to maximize deviations, and a realistic astrophysical population. Biases are assessed in the population at the level of single-event binary parameters, as well as hierarchical EOS constraints. The qURs we choose to investigate are:
\begin{enumerate}[label=(\roman*)]
    \item \textit{Spin induced quadrupole moment} (SIQM): $\Lambda \rightarrow C_Q$ where $C_Q$ is the spin-induced deformability which translates to the SIQM $Q=C_Qm^3\chi^2$ with NS mass $m$ and dimensionless aligned spin $\chi$.
    \item \textit{f-mode frequency}: $\Lambda \rightarrow f_2$ where $f_2$ is the $f$-mode oscillation frequency of the NS.
    \item \textit{Binary Love}: $(\Lambda_s, q)\rightarrow\Lambda_a$ where the anti-symmetric $\Lambda_a= \frac{1}{2}(\Lambda_2 - \Lambda_1)$ and symmetric $\Lambda_s= \frac{1}{2}(\Lambda_2 + \Lambda_1)$ tidal combinations are connected alongside the mass ratio $q=m_1/m_2\geq1$.
\end{enumerate}

The paper is organised as follows. In Sec.~\ref{sec:EOSs} we outline the construction of the EOSs used in this study. In Sec.~\ref{sec:methods} we describe the employed methods, including a summary of Bayesian inference, the parameter estimation setup, the mismatch procedure, the population generation and hierarchical population inference. We examine the effect of the SIQM, \textit{f}-mode frequency, and binary Love qUR on single-event posteriors in Secs.~\ref{sec:SIQM}, \ref{sec:fmode}, and \ref{sec:BL}, respectively. In Sec.~\ref{sec:pop} we study their impact on hierarchical EOS constraints. Finally we conclude in Sec.~\ref{sec:conclusions}.
Throughout the work, we set $G=c=1$. For the individual component masses we define $m_1$ and $m_2$ as the primary and secondary masses respectively, where these subscripts carry forward onto other NS parameters. The mass ratio is taken as $q=m_1/m_2\geq1$, and aligned spin components are given by $\chi_1$ and $\chi_2$.




\section{Equations of State}
\label{sec:EOSs}
\begin{figure}
    \centering
    \includegraphics[width=\columnwidth]{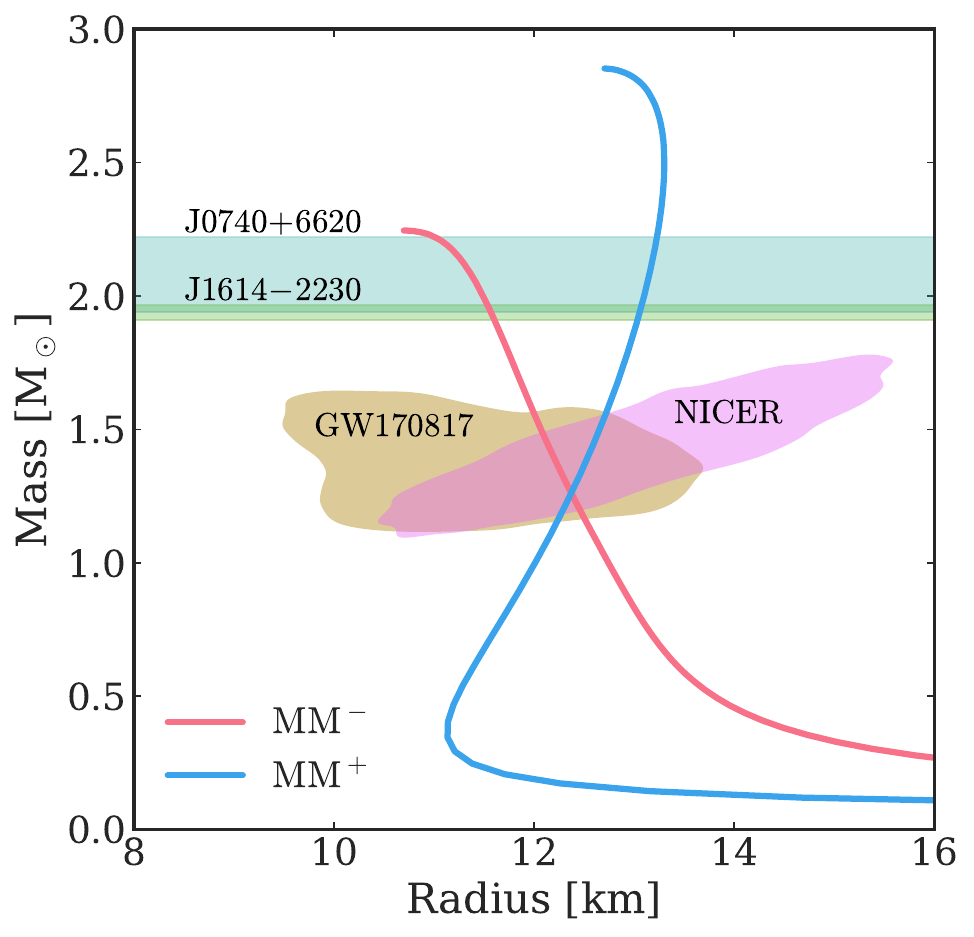}
    \captionsetup{justification=raggedright,singlelinecheck=false}
    \caption{Mass-radius curves for MM$^{-}$ (pink) and MM$^{+}$ (blue) with nuclear parameters listed in App. \ref{Appdx:mm}. The observed masses to 2$\sigma$ of the millisecond pulsar PSR J0740+6620~\cite{Fonseca:2021wxt} (turquoise) and PSR J1614-2230~\cite{NANOGrav:2023hde} (green) are shown alongside 90\% confidence constraints from the analysis of GW170817~\cite{LIGOScientific:2017vwq, LIGOScientific:2018cki, LIGOScientific:2018hze} (yellow) and observations reported by the NICER mission for PSR J0030+0451~\cite{Miller:2019cac} (light pink).}
    \label{fig:EOScurves}
\end{figure}

We consider two deliberately `extreme' EOSs constructed within a meta-model (MM)~\cite{Margueron2018a, Grams22a} framework, denoted as MM$^{-}$ and MM$^{+}$, where the superscript refers to the respective slope of the mass–radius relation . These are shown in Fig.~\ref{fig:EOScurves}. Following the strategy of Ref.~\cite{Raithel:2022orm}, these EOSs are designed to probe regions where deviations from qURs are expected to be enhanced. Despite their extremity, both models satisfy various astrophysical constraints: they are consistent with the $90\%$ confidence intervals inferred from GW170817~\cite{LIGOScientific:2017vwq, LIGOScientific:2018cki, LIGOScientific:2018hze}, with the NICER radius measurements for PSR J0030+0451 and J0740+6620~\cite{Miller:2019cac, Riley:2019yda, Miller:2021qha, Riley:2021pdl}, and with the observed high-mass pulsars PSR J0740+6620 and PSR J1614–2230~\cite{Fonseca:2021wxt, NANOGrav:2023hde}. By adopting such contrasting EOSs rather than a canonical model, we aim to test the robustness of qURs under conditions where they are most likely to break down.

We compute unified EOSs in which the crust and core are consistently obtained using a compressible liquid drop model, as detailed in Ref.~\cite{Grams22a}. The nuclear MM~\cite{Margueron2018a} expresses the potential energy per particle as a Taylor expansion around saturation density, linking the energy density functional to nuclear experiments and ab initio theory in a flexible, semi-agnostic approach. 
Such flexibility allows the MM to reproduce a wide variety of existing functionals~\cite{Dinh23, Grams22b}, or to efficiently explore extreme regions of the parameter space and generate large ensembles of EOSs for Bayesian analyses~\cite{Margueron18b, Mondal23}. In the present work, we exploit this property to generate two EOSs with maximally distinct mass–radius slopes. Further details of the EOS construction are provided in Appendix~\ref{Appdx:mm}.

\section{Methodology}
\label{sec:methods}
We explore the effect of using qURs in measuring NS properties through Bayesian parameter estimation.
The posterior density distribution function of a set of model parameters $\boldsymbol{\theta}$ given the data $d$ is 
\begin{align}
    p(\boldsymbol{\theta}|d) &= \frac{ \mathcal{L}(d|\boldsymbol{\theta}) \pi(\boldsymbol{\theta})}{\mathcal{Z}_d},
\end{align}
where $\mathcal{L}(d|\boldsymbol{\theta})$ denotes the likelihood, $\pi(\boldsymbol{\theta})$ the prior, and $\mathcal{Z}_d$ the evidence or marginalised likelihood, defined as
\begin{equation}
    \mathcal{Z}_d = \int \mathcal{L}(d|\boldsymbol{\theta}) \pi (\boldsymbol{\theta}) d\boldsymbol{\theta}.
\end{equation}
The set of parameters $\boldsymbol{\theta}$ that we consider include the intrinsic binary parameters such as the component masses, the dimensionless spin of the components, and the quadrupolar tidal deformabilites, and extrinsic parameters, i.e., the sky location, inclination, distance, polarisation, coalescence time and phase. We obtain one-dimensional and two-dimensional posteriors by marginalising over other parameters. 

Throughout this study, we model the GW signal with the \textsc{IMRPhenomXAS\_NRTidalv3} waveform model~\cite{Pratten:2020fqn, Abac:2023ujg}, which accounts for unequal-mass, aligned-spin BNS systems and incorporates dynamical tidal effects. Parameter estimation is performed throughout using the \textsc{Bilby} inference library~\cite{Ashton:2018jfp} in conjunction with the nested sampler \textsc{Dynesty}~\cite{Speagle:2019ivv}. We choose a configuration of two 15km L-shaped Einstein Telescope detectors placed in Limburg and Sardinia respectively with power spectral densities as in Ref.~\cite{et_psd}. We use a minimum frequency of 5Hz, a maximum of 2048Hz, and a sampling rate of 4096Hz in all analyses. For parameter estimation acceleration, we apply a multibanded likelihood~\cite{Vinciguerra:2017ngf,Morisaki:2021ngj} and marginalise over the luminosity distance $D_L$. We sample in the chirp mass $\mathcal{M}_c$, mass ratio $q$, aligned component spins $\chi_1$ and $\chi_2$, tidal parameters $\tilde{\Lambda}$ and $\delta\tilde{\Lambda}$~\cite{Wade:2014vqa}, source-frame luminosity distance $D_L$, and sky-location and orientation angles, which are assumed to be isotropically distributed. Priors are uniform in component masses, aligned spins, $\tilde{\Lambda}$, $\delta\tilde{\Lambda}$, and $D_L$ in comoving volume and source frame time, with ranges $\mathcal{M}_c \in [0.7, 2.0] M_\odot$, $q \in [0.4, 1]$, $\chi_1, \chi_2 \in [-0.15, 0.15]$, $D_L \in [0, 1200] \mathrm{Mpc}$, $\tilde{\Lambda} \in [0, 5000]$, and $\delta\tilde{\Lambda} \in [-5000, 5000]$.

\subsection{Selected  Cases}
\label{subsec:extreme_cases_method}
For each qUR, we first perform parameter estimation on selected binaries chosen to maximise the disagreement between the waveform using the qUR, $h^{\mathrm{UR}}$, and the waveform using the NS properties computed from the EOS, $h^{\mathrm{EOS}}$. We quantify the degree of disagreement through the mismatch:
\begin{equation}
\label{eq:MM}
    \mathcal{MM}(h^{\mathrm{UR}}, h^{\mathrm{EOS}}) = 1 - \mathrm{max}_{\phi_c, t_c} \frac{\langle h^{\mathrm{UR}}(\phi_c, t_c)|h^{\mathrm{EOS}}\rangle}{\sqrt{\langle h^{\mathrm{UR}}|h^{\mathrm{UR}}\rangle\langle h^{\mathrm{EOS}}|h^{\mathrm{EOS}}\rangle}},
\end{equation}
with the noise weighted inner product being
\begin{equation}
    \langle h^{\mathrm{UR}}|h^{\mathrm{EOS}}\rangle = 4 
  \ \mathrm{Re}\int^{f_{\mathrm{max}}}_{f_{\mathrm{min}}}\frac{\tilde{h}^{\mathrm{UR}}\tilde{h}^{\mathrm{EOS}}}{S_n}df,
\end{equation}
where the mismatch is minimised over an arbitrary coalescence phase $\phi_c$ and time $t_c$ shift, and $\tilde{h}$ denotes the Fourier transform $h(f)$ of $h(t)$. $S_n$ is the power spectral density of the detector, which we set to $S_n=1$ such that the mismatch is detector agnostic.\footnote{For each qUR, we compute waveform mismatches between $f_{\rm min} =  5$ Hz and $f_{\rm max} = 2048 $ Hz across a uniform 200$\times$200 grid of component masses $m \in [1.0, 2.2],M_\odot$ which determines $\Lambda$ for the given EOS, holding all parameters not connected to the qUR fixed so that any differences arise solely from the qUR deviation. Further details are given in Appendix \ref{Appdx:mismatches}.}

We then simulate signals with parameters corresponding to the maximum mismatch point computed on the grid, and parameter estimation. We systematically vary whether qURs are used in the injection, the recovery, or both depending on the qUR being studied. This allows us to assess the largest potential biases each qUR can introduce, giving a worst-case estimate. These parameter estimation runs are performed with zero noise, so that any observed biases arise solely from the qURs and not from random noise fluctuations. We adjust the distances of the sources to yield an SNR = 217.97, which corresponds to a 1$M_{\odot}$+1$M_{\odot}$ system placed at 70Mpc for our chosen sky location. 

\subsection{Population}
\label{subsec:population}
\subsubsection{Population Generation}
We then perform parameter estimation on an astrophysically motivated population of sources. For this population we simulate 1000 binaries with component masses drawn from the \textsc{FLAT\_Q} model distribution~\cite{Landry:2021hvl} in interval $m\in[1,2.2] \, M_{\odot}$ to span the mass range common to both EOSs, aligned spins drawn uniformly $\chi\in[-0.15, 0.15]$ to reflect the low expected spins of NSs from the observed population~\cite{Burgay:2003jj}, luminosity distance drawn uniformly in comoving volume $D_L\in[10,1000]$Mpc, and isotropically distributed sky location. The tidal deformabilities for each binary are computed from the component masses according to each EOS. From this population we then select the 20 events with the largest SNRs (ranging 195-668) to use in this study. This number of high-SNR BNS events is expected to be detected by the ET configuration employed in roughly one year of observation~\cite{Branchesi:2023mws}.

\subsubsection{Hierarchical Inference}
For these 20 events, we perform hierarchical EOS inference to assess whether biases arise at the population level when adopting qURs following the method of Ref.~\cite{Lackey:2014fwa}. Let $D=\{d_1,\dots,d_n\}$ denote the data from the $n=20$ loudest events in the simulated population, and $\theta_i$ denote all parameters describing the $i$-th event. The posterior on EOS parameters $\vec{E}$ can be written as 
\begin{equation}
    p(\vec{E}|D)=\int\prod^n_{i=1}d\theta_i \, p(\vec{E}|\mathcal{M}^{\rm src}_{c, i}, q_i, \tilde{\Lambda}_i,  \delta\tilde{\Lambda}_i)\,p(\theta_i|d_i)
\end{equation}
where $\mathcal{M}^{\rm src}_{c}$ is the source frame chirp mass. For a given EOS, the tidal deformabilites ($\tilde{\Lambda}$, $\delta\tilde{\Lambda}$) are fully determined by the component masses which can be computed from ($\mathcal{M}^{\rm src}_{c}$, $q$). Thus the conditional EOS posterior for a single event is
\begin{equation}
\begin{aligned}
    p(\vec{E}|&\mathcal{M}^{\rm src}_{c, i}, q_i, \tilde{\Lambda}_i,  \delta\tilde{\Lambda}_i) = p(\vec{E})\, p(\mathcal{M}^{\rm src}_{c, i}, q_i|\vec{E}) 
    \\
    &\times \delta(\tilde{\Lambda}_i - \tilde{\Lambda}(\vec{E}, \mathcal{M}^{\rm src}_{c, i}, q_i)) \, \delta(\delta\tilde{\Lambda}_i - \delta\tilde{\Lambda}(\vec{E}, \mathcal{M}^{\rm src}_{c, i}, q_i))\,.
\end{aligned}
\end{equation}
Here, the conditional prior $p(\mathcal{M}_c^{\rm src},q|\vec{E})$ enforces the mass range supported by the EOS. For each event, the likelihood marginalized over extrinsic and nuisance parameters $\theta_{\rm ex}$ is
\begin{equation}
\begin{aligned}
    \mathcal{L}_i(d_i| \mathcal{M}_{c, i}&, q_i, \tilde{\Lambda}_i, \delta\tilde{\Lambda}_i) =
    \int d\theta_{\mathrm{ex}}\,p(\theta_{\mathrm{ex}, i}) \\
    &\times p(d_i|\mathcal{M}_{c, i}, q, i, \tilde{\Lambda}, i, \delta\tilde{\Lambda}, i, \theta_{\mathrm{ex},  i}),
\end{aligned}
\end{equation}
which represents the likelihood of the data given only the relevant parameters for the EOS.

In practice, we approximate this quasilikelihood using a four-dimensional Gaussian kernel density estimate constructed from posterior samples of $(\mathcal{M}_{c,}, q, \tilde{\Lambda}, \delta\tilde{\Lambda})$ for each event. 

The EOS-level posterior is then obtained by marginalising over the masses:
\begin{equation}
\begin{aligned}
    p(\vec{E}|D) &\propto \prod_{i=1}^n \int d\mathcal{M}^{\rm src}_{c, i}\,dq_i \  p(\vec{E}|\mathcal{M}^{\rm src}_{c, i}, q_i, \tilde{\Lambda}_i,  \delta\tilde{\Lambda}_i)\\
    &\times \mathcal{L}_i(d_i\,\big|\,\mathcal{M}_{c, i},q_i,
    \tilde{\Lambda}_i,
    \delta\tilde{\Lambda}_i).
    \label{eq:EOSposterior}
\end{aligned}
\end{equation}
This posterior is evaluated using nested sampler \textsc{Dynesty}, and reconstruct the mass–tidal deformability relation $\Lambda(m)$ from the inferred EOS parameters.

We construct our reference EOSs using the nuclear MM, a phenomenological energy-density functional directly tied to empirical nuclear-matter parameters and to the microphysical composition of dense matter.
For inference, we use the spectral EOS parameterisation~\cite{Lindblom:2010bb} as implemented within \textsc{LalSimulation}~\cite{lalsuite}. In the spectral decomposition, the adiabatic index $\Gamma$ of the EOS with pressure $p$ and energy density $\epsilon$, is defined by
\begin{equation}
    \Gamma(p) = \frac{\epsilon+p}{p}\frac{dp}{d\epsilon} ,
\end{equation}
which is spectrally composed onto a set of polynomial basis functions
\begin{equation}
\Gamma(y) = \exp \bigg(\sum_k\gamma_ky^k\bigg)
\end{equation}
where $y=\log (p/p_0)$ is a dimensionless pressure relative to the reference pressure $p_0$. $\gamma_k$ are the expansion coefficients. This description ensures that there are no discontinuities in the derivative of the EOS. We use a four-parameter spectral representation $\vec{E}=\{\gamma_1, \gamma_2, \gamma_3, \gamma_4\}$ with uniform priors in the intervals $\gamma_1\in [0.2,2]$, $\gamma_2\in [-1.6,1.7]$, $\gamma_2\in [-0.6,0.6]$ and $\gamma_4\in [-0.02,0.02]$. Within the sampling, we use the principle component analysis decomposition of the spectral coefficients described in~\cite{Wysocki:2020myz}, and enforce EOS causality.

\section{Spin Induced Quadrupole moment}  
\label{sec:SIQM}
\subsection{Implementation}
Rotation of NSs results in additional deformation, producing SIQMs and octopole moments, parameterised by $C_{\rm Q}$ and $C_{\rm Oct}$, respectively.  This effect is incorporated within the phase of the \textsc{NRTidalv3} model through post Newtonian (PN) self-spin corrections up to 3.5PN~\cite{Dietrich:2019kaq, Marsat:2014xea,Bohe:2015ana}, where the full expressions can be found in \cite{Abac:2023ujg}. For a black hole $C_{\rm Q} = C_{\rm Oct} = 1$, however, for NSs these are EOS-dependent quantities with $C_{\rm Q} \neq 1$ and $C_{\rm Oct} \neq 1$. Within \textsc{NRTidalv3} the SIQM is computed for each NS with the qUR~\cite{Yagi:2016bkt}
\begin{equation}
    \begin{split}
        \log(C_{\rm Q}) =& 0.1940 + 0.0916\log(\Lambda)+0.04812\log^2(\Lambda) \\ &-0.0004286\log^3(\Lambda)+0.00012450\log^4(\Lambda) \, . \\
    \end{split}
    \label{eq:CQ}
\end{equation}
The corresponding qUR for $C_{\rm Oct}$ is given in \cite{Abac:2023ujg}, which we utilise throughout this work to focus only on the leading order quadrupole effect of $C_{\rm Q}$.

\begin{figure}[htbp!]
    \centering
        \includegraphics[width=\columnwidth]{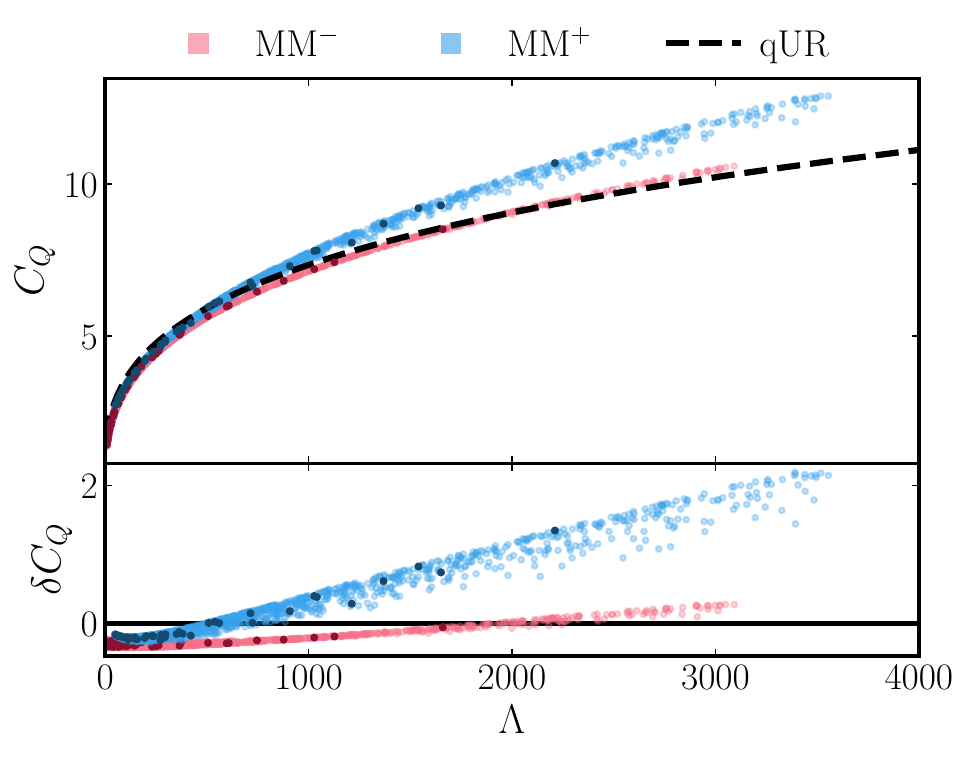}
    \captionsetup{justification=raggedright,singlelinecheck=false}
    \caption{\textit{Spin-induced quadrupole moment}: Comparison of the qUR for the SIQM (black dashed lines) for MM$^{-}$ (pink) and MM$^{+}$ (blue). Points show a population of 1000 sources, with the 20 highest-SNR sources highlighted. The top panel displays the qUR for $C_Q$, while bottom panel shows deviations from the qUR, i.e., $\delta C_Q\equiv C_Q^{\rm EOS}-C_Q^{\rm qUR}$.}
    \label{fig:QUR}
\end{figure}

Within our analyses, we use the default implementation of the qUR, but also introduce code modifications to the model publicly available in LALsuite~\cite{lalsuite} to treat the quadrupole moment as a free parameter, which allows us to set it explicitly and sample $C_Q$ independently. We calculate the SIQM for our waveform injections using a fit calibrated to results from the RNS code~\cite{Stergioulas:1994ea, Stergioulas:2003yp}, which models uniformly rotating neutron stars, and extrapolated to the low-spin regime (see Appendix~\ref{Appdx:SIQM}).

Figure~\ref{fig:QUR} compares the qUR with the SIQMs calculated from the EOSs for the sources in our population catalog. We observe MM$^{-}$ closely follows the qUR, whereas MM$^{+}$ deviates, with the qUR systematically underestimating the SIQM, particularly as the tidal deformability increases.

\subsection{Selected Cases}
We first perform parameter estimation on systems designed to exhibit the largest deviations from the qUR. We select these systems by maximising the mismatch (Eq.~\ref{eq:MM}) between waveforms $h^{\rm UR}$ which use the qUR for $C_Q$ (Eq.~\ref{eq:CQ}), and waveforms $h^{\rm EOS}$ which compute $C_Q$ from the EOS though the RNS fit. We fix the spins of both NSs at $\chi_A=0.15$, as the largest value that we consider. The mismatches for  each EOS for all qURs are shown in Appendix~\ref{Appdx:mismatches}. The source masses for the point of maximum deviation from the qUR are $(m_1, m_2)=(1.35M_\odot, 1.35M_\odot)$ for MM$^{-}$ and $(m_1, m_2)=(1M_\odot, 1M_\odot)$ for MM$^{+}$.

We perform three sets of analyses for each of these configurations:
\begin{itemize}
\item \textit{Baseline}: We simulate the waveform using the qUR for $C_Q$ as implemented in \textsc{NRTidalv3}, and perform parameter estimation assuming the same qUR. This serves as a reference point for comparison.
\item \textit{Universal Relation}: We simulate the waveform using the values obtained from the EOS SIQM fits, and perform parameter estimation assuming the qUR for $C_Q$. This allows us to quantify biases arising when using a qUR that significantly deviates from the true values.
\item \textit{$C_Q$ Sampling}: We simulate the waveform using the values obtained from the EOS SIQM fits, and perform parameter estimation independently sampling $C_Q$ for each NS. This provides an assessment of the impact of treating $C_Q$ as a free parameter in the inference. The additional prior on $C_Q$ in this analysis is uniform over interval $C_Q\in[0,20]$.
\end{itemize}

\begin{figure*}[htbp!]
    \centering
        \centering
        \includegraphics[width=\textwidth]{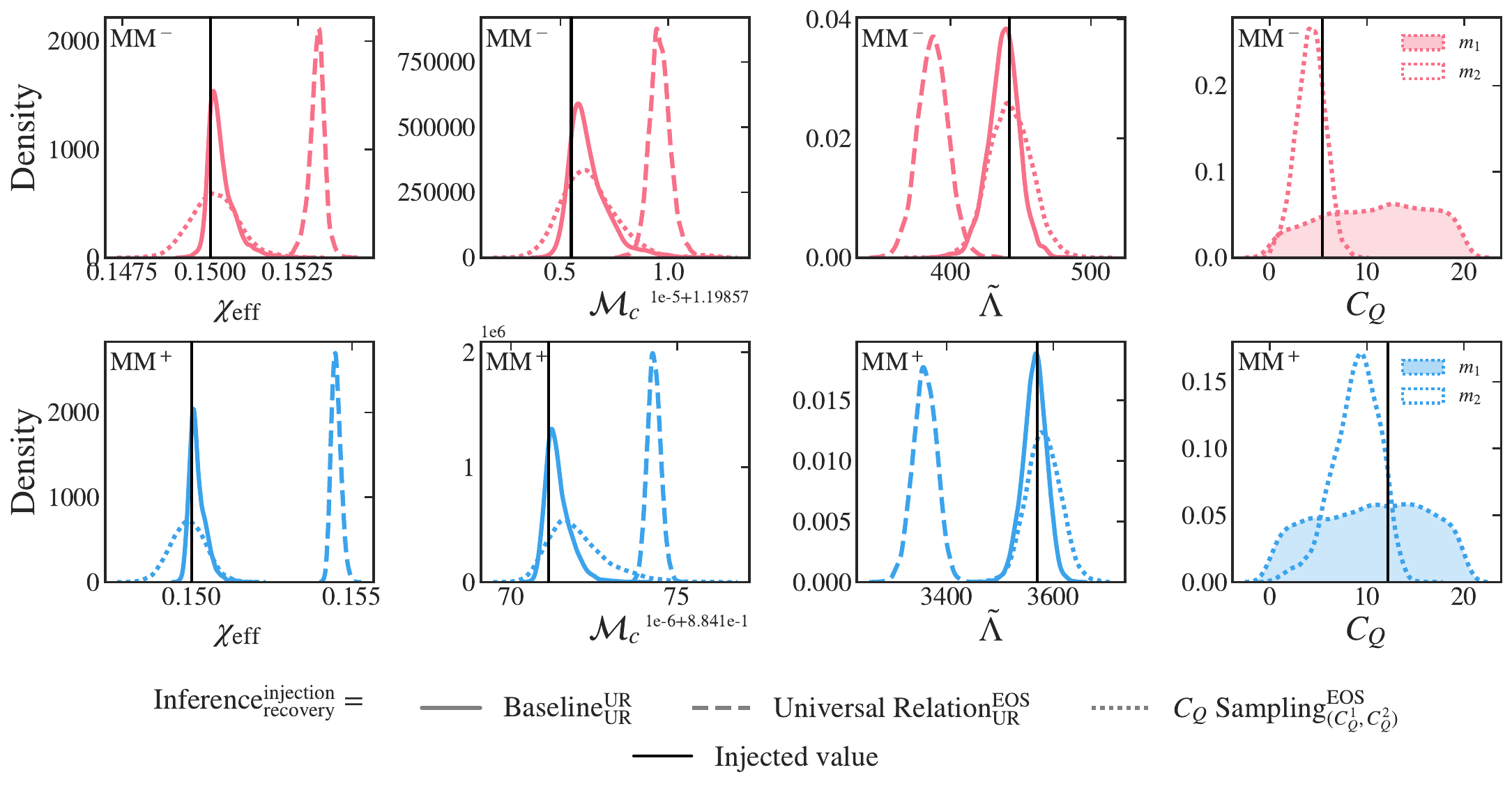}\captionsetup{justification=raggedright,singlelinecheck=false}
        \caption{\textit{Spin-induced quadrupole moment:} Posterior density distributions when simulating signals with EOSs MM$^-$ (pink) and MM$^+$ (blue) for the Baseline (solid), Universal Relation (dashed), and $C_Q$ sampling (dotted) analyses as described in the text. True injected values are indicated (black solid lines). The posteriors shown are \textit{First column}: Effective spin $\chi_{\rm eff}$. \textit{Second column}: Detector frame chirp mass $\mathcal{M}_c$. \textit{Third column}: Joint tidal deformability $\tilde{\Lambda}$. \textit{Fourth column}: SIQM $C_Q$ in the case of the $C_Q$ sampling analysis, for the components $m_1$ (shaded) and $m_2$ (unshaded). Injected values for each component mass overlap as these are equal-mass, equal-spin systems.}
        \label{fig:Qpost}
    \label{fig:extremeQpole}
\end{figure*}

We show the recovered posteriors from these analyses in Fig.~\ref{fig:Qpost}. We choose to show the effective binary spin $\chi_{\mathrm{eff}}$, a mass weighted combination of the individual spins, which for aligned spin systems is
\begin{equation}
  \chi_{\rm eff}  = \frac{m_1\chi_1+m_2\chi_2}{m_1+m_2} \, .
\end{equation}

We find that when the SIQM qUR is consistently used in both the simulation and the analysis, or when the waveform is generated using EOS-derived SIQM while the analysis treats $C_Q$ as a free parameter, the intrinsic parameters are accurately recovered. As expected, the qUR-based recovery yields narrower posteriors due to the reduced dimensionality of the parameter space. In contrast, enforcing the qUR in the analysis when the true SIQM is set by the EOS introduces significant biases in multiple parameters. For both EOSs considered, the inferred effective spin and chirp mass shift toward larger values, while the tidal deformability is biased toward smaller values, with the injected values confidently excluded from the posteriors. In the $Q$-sampling analysis, we recover $C_Q$ for the secondary component, while the primary remains prior-dominated. This indicates that third-generation detectors will enable direct measurements of $C_Q$ for sufficiently rapidly rotating NSs and indicated by~\cite{Samajdar:2020xrd,Zheng:2025nij}.

\subsection{Population Draws}
Thus far we have investigated the biases arising from the SIQM qUR in systems specifically constructed to maximise them. Next we examine the effect on a population of binaries, both to characterise the biases expected in more canonical systems, and to determine their influence on EOS measurements obtained from a population. We therefore conduct parameter estimation on our population of the 20 high-SNR events as explained in Sec.~\ref{subsec:population}. We use identical parameter estimation settings in the analysis as in the selected cases for all events, similarly with zero noise. 
\begin{figure*}[ht!]
    \centering
    \includegraphics[width=0.8\linewidth]{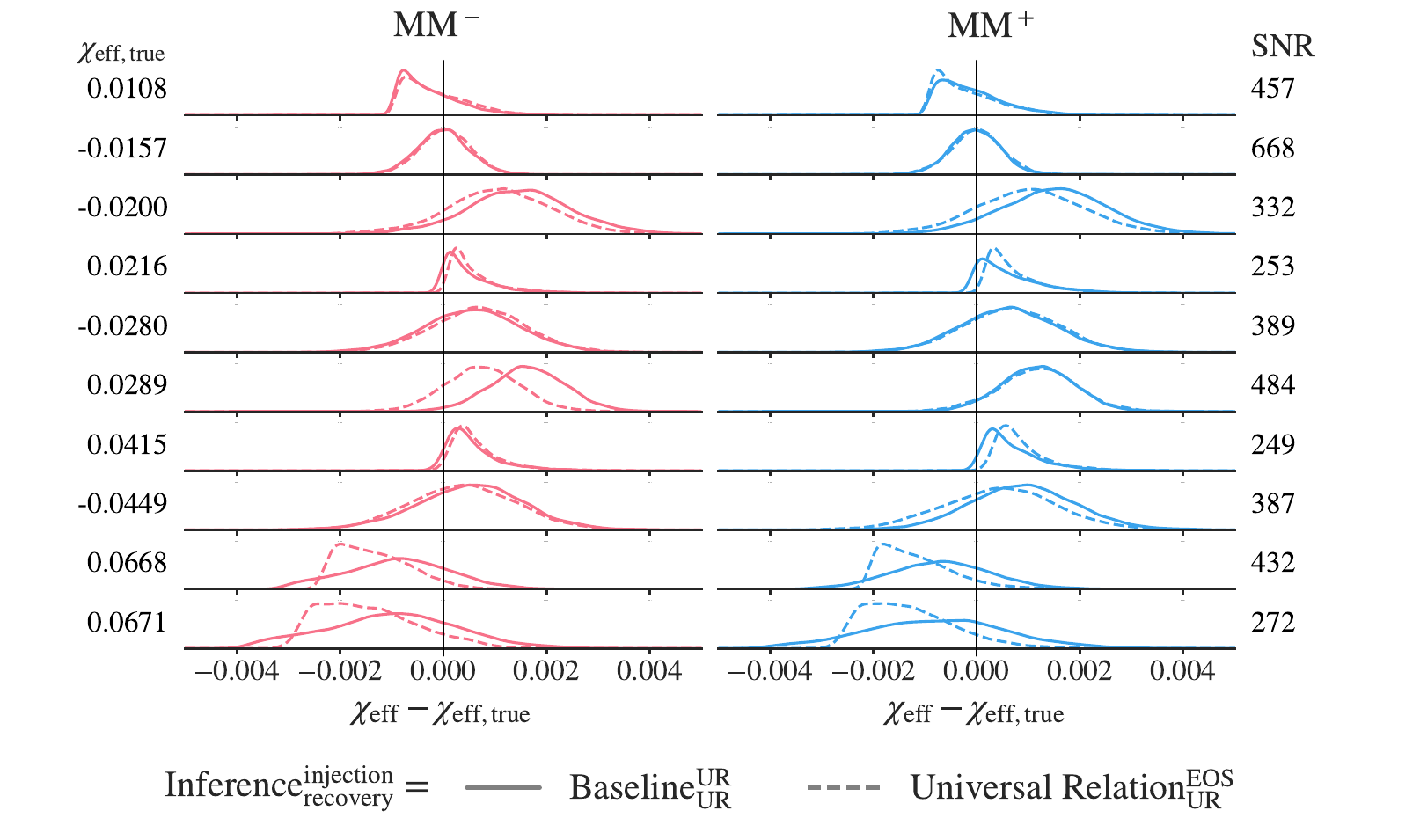}
\captionsetup{justification=raggedright,singlelinecheck=false}
    \caption{\textit{Spin-induced quadrupole moment:} Posterior probability distributions of $\chi_{\mathrm{eff}} - \chi_{\mathrm{eff,true}}$, where $\chi_{\mathrm{eff,true}}$ is the true injected effective spin, for MM$^{-}$ (left) and MM$^{+}$ (right) EOS. An unbiased measurement corresponds to zero (black line). Results are shown for the ten highest-SNR events, ordered by the magnitude of the injected $\chi_{\mathrm{eff, true}}$, with the SNRs listed on the right. Two analyses are compared as described in the text: Baseline (solid) and Universal Relation (dashed). We choose to omit the $C_Q$ sampling analysis for clarity.}
    \label{fig:Qpop}
\end{figure*}
A representative sample of the posteriors from the ten highest SNR sources is given in Fig.~\ref{fig:Qpop}, showing the posteriors of $\chi_{\mathrm{eff}}$ relative to the true injected value $\chi_{\mathrm{eff,true}}$. This sample shows weak evidence for larger biases at higher spin magnitudes, though the trend is not monotonic. Several systems even exhibit smaller deviations than the case where both injection and recovery use the qUR. This reflects the fact that bias depends not only on spin, but also on other binary properties such as masses and orientations. For clarity, the results of the runs in which the quadrupole moment is sampled independently are not included in Fig.~\ref{fig:Qpop}. These analyses reproduce the features seen in the selected cases, showing consistency with the baseline run but with larger uncertainties. For most events, where the NS spins are low, the posteriors for the SIQM are prior-dominated, indicating that the data provide essentially no constraint on $C_Q$. Meaningful recovery of $C_Q$ occurs only in systems with sufficiently high spin, reproducing the trends seen in the selected cases. 

\section{Fundamental Mode Frequency}
\label{sec:fmode}
\subsection{Implementation}
The tidal deformability $\Lambda$ typically represents the leading-order \textit{adiabatic} tides, where the quadrupolar response of the NS is assumed instantaneous with the tidal field generated by its companion. This assumption holds in the limit that the GW frequency is much smaller than the frequency of the fundamental oscillation mode of the NS $f_2$, however in the late inspiral this assumption can break down, producing the so-called \textit{dynamical} tidal effects~\cite{Steinhoff:2016rfi, Lai:2006pr, Hinderer:2016eia, Steinhoff:2021dsn} for which the tidal deformability becomes time or frequency dependent. In the third-generation era, it is vital that dynamical effects be accurately included within waveform models to avoid waveform systematics~\cite{Pratten:2021pro, Bretz:2026asa}.

The qUR for the $f$-mode frequency rescaled to a 1.4$M_\odot$ NS $M_{1.4}\, f_2=M\,/\,1.4M_{\odot} \, f_2$, is given by~\cite{Sotani:2021kiw}
\begin{equation}
    M_{1.4}\, f_2\, [\mathrm{kHz}] = \sum_{i=0}^5 g_i (\log_{10}\Lambda)^i
    \label{eq:fmodeUR}
\end{equation}
where
\begin{align}
    g_i \in [&4.2590, -0.47874, -0.45353, \nonumber \\
    & 0.14439, -0.016194, 0.00064163] \, .
\end{align}
The effect of dynamical tides can be incorporated with a time-dependent enhancement factor~\cite{Steinhoff:2016rfi}, which is transformed into the frequency domain using the stationary phase approximation and requires solving a second order differential equation. As this is a computationally expensive process,  in \textsc{NRTidalv3} the frequency-domain dynamical tidal contribution is instead approximated with a phenomenological fit\footnote{We also note that in \textsc{NRTidalv3}, dynamical tides are modelled assuming non-rotating NSs. The intrinsic spin of the individual stars can shift the $f$-mode frequency, which affects the waveform phase. At the moment, this effect is not included in the \textsc{NRTidalv3} model.} calibrated to 10 EOSs across 55 binary configurations of the numerical relativity data used within the model. Although this approach increases efficiency, it means that it is not possible to freely sample the $f$-mode frequency within \textsc{NRTidalv3} as currently implemented. To address this, we recalibrate the frequency-domain phenomenological fit for the f-mode frequency used in the \textsc{NRTidalv3} for each of our specific EOSs, using the $f$-mode frequencies determined in two ways: (a) directly from the EOS, and (b) from the qUR. Comparing these two implementations allows us to isolate the effect of the qUR itself, independent of the waveform systematic uncertainty introduced by the fit. However, we point out that we do not recalibrate the entire \textsc{NRTidalv3} model, i.e., only modify one small ingredient within the entire calibration process described in~\cite{Abac:2023ujg}. Nevertheless, our approach is necessary to quantify the direct impact of the qUR and the waveform systematic error arising from the calibration procedure. We stress that \textsc{NRTidalv3} performs well for a broad range of EOSs, and the waveform systematic in our study arises from using `extreme` EOSs outside its calibration set. The details of our procedure, and the associated waveform systematics when using a differing dataset, are provided in App.~\ref{Appdx:dynamicaltides}.
\begin{figure}[htbp!]
    \centering
        \includegraphics[width=\columnwidth]{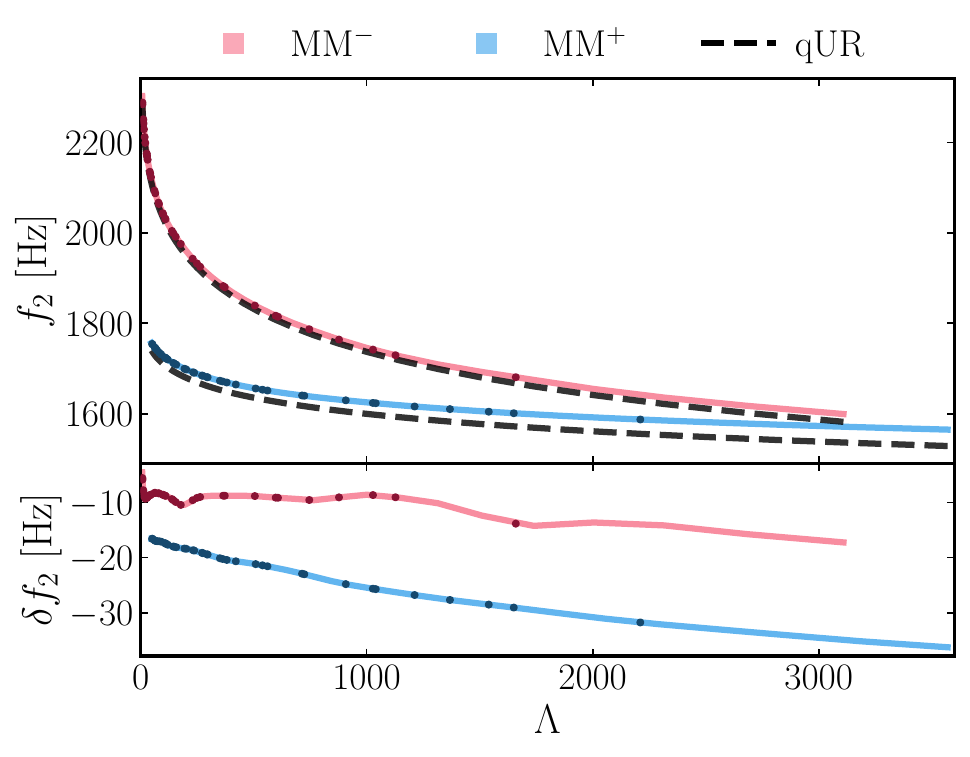}
    \captionsetup{justification=raggedright,singlelinecheck=false}
    \caption{\textit{Fundamental mode frequency:} $f$-mode frequency $f_2$ obtained from MM$^{-}$ (pink) and MM$^{+}$ (blue) between masses of $1M_\odot$ and $2.2M_\odot$. Highlighted points show the components from the $20$ loudest binary sources of the population. \textit{Top panel}: absolute $f$-mode frequencies compared with the qUR reference (black dashed line). \textit{Bottom panel}: Deviations of the frequencies from the qUR, i.e., $\delta f_2\equiv f_2^{\rm qUR}-f_2^{\rm EOS}$.}
    \label{fig:fUR}
\end{figure}

Figure~\ref{fig:fUR} shows the qUR for the $f$-mode frequency along with the values obtained for each EOS (details on how this is calculated are given in App.~\ref{Appdx:fmode}). We show the qUR in physical units here (Hz) for clarity rather than the rescaled version of Eq.~\ref{eq:fmodeUR}, hence the appearance of a distinct qUR for each EOS. Across both EOSs, the qUR systematically underestimates the $f$-mode frequency. Again, MM$^{+}$ exhibits larger deviations from the qUR than MM$^{-}$, and both EOS show deviations of $\mathcal{O}(10)$ Hz.

\subsection{Selected Cases}
We find the configurations that maximise the deviation from the qUR at $(m_1, m_2)=(1.22M_\odot, 1.10M_\odot)$ for MM$^{-}$ and $(m_1, m_2)=(1.09M_\odot, 1.08M_\odot)$ for MM$^{+}$ as detailed in Appendix~\ref{Appdx:mismatches}

\begin{figure}[htbp!]
    \centering
        \centering
        \includegraphics[width=\columnwidth]{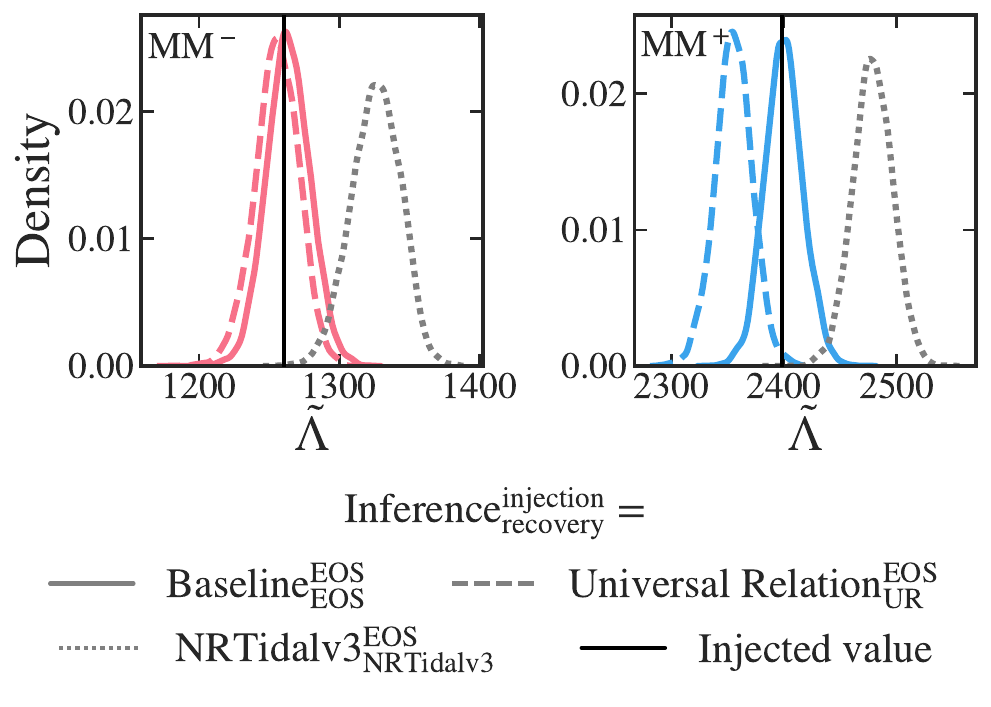}
        \captionsetup{justification=raggedright,singlelinecheck=false}
        \caption{\textit{Fundamental mode frequency:} Posterior density distributions of $\tilde{\Lambda}$ for the selected $f$-mode frequency case with EOSs MM$^{-}$ (left) and MM$^{+}$ (right) for three analyses described in the text: Baseline (solid), Universal Relation (dashed) and NRTidalv3 (dotted). True injected values are indicated (vertical black lines).}
        \label{fig:fpost}

\end{figure}

We perform three sets of analyses for each of these configurations:
\begin{itemize}
\item \textit{Baseline}: We simulate and analyse signals using \textsc{NRTidalv3} with our enhancement factor fit recalibration using the $f$-mode frequencies directly computed from the EOS.
\item \textit{Universal Relation}: We simulate the signal using \textsc{NRTidalv3} with our recalibration using EOS $f$-mode frequencies, and perform parameter estimation using \textsc{NRTidalv3} with the our recalibration using $f-$mode frequencies computed through the qUR. This allows us to quantify biases arising from the qUR.
\item \textit{NRTidalv3}: We simulate the signal with our recalibrated \textsc{NRTidalv3} using EOS $f$-mode frequencies (identical to the baseline), and perform parameter estimation using the original \textsc{NRTidalv3} fit. This demonstrates bias from waveform systematics that arise due to the phenomenological fitting process and \textit{not} solely from the qUR.
\end{itemize}

As noted previously, the $f$-mode frequencies cannot be sampled freely due to the frequency-domain implementation of \textsc{NRTidalv3}. 

The recovered posteriors for the leading order tides $\tilde{\Lambda}$ are shown in Fig.~\ref{fig:fpost}.
Consistent with expectations, the baseline shows that the injected $\tilde{\Lambda}$ is recovered. When the analysis assumes the qUR we observe a small bias in the case of MM$^-$ and a moderate bias for MM$^+$. The bias pushes towards a softer EOS, which is expected since the qUR systematically underestimates the $f$-mode frequency which in turn artificially enhances the dynamical tidal effects. Thus to compensate, the tidal deformability is biased to lower values. We also show the posterior from the analysis with the original \textsc{NRTidalv3} model, which shows significant biases towards a stiffer EOS for both MM$^-$ and MM$^+$. This follows from the introduction of a waveform systematic error associated with the frequency-domain dynamical tides fitting procedure in \textsc{NRTidalv3} (see Appendix \ref{Appdx:dynamicaltides}).

\begin{figure*}[htpb]
    \centering
    \includegraphics[width=0.9\linewidth]{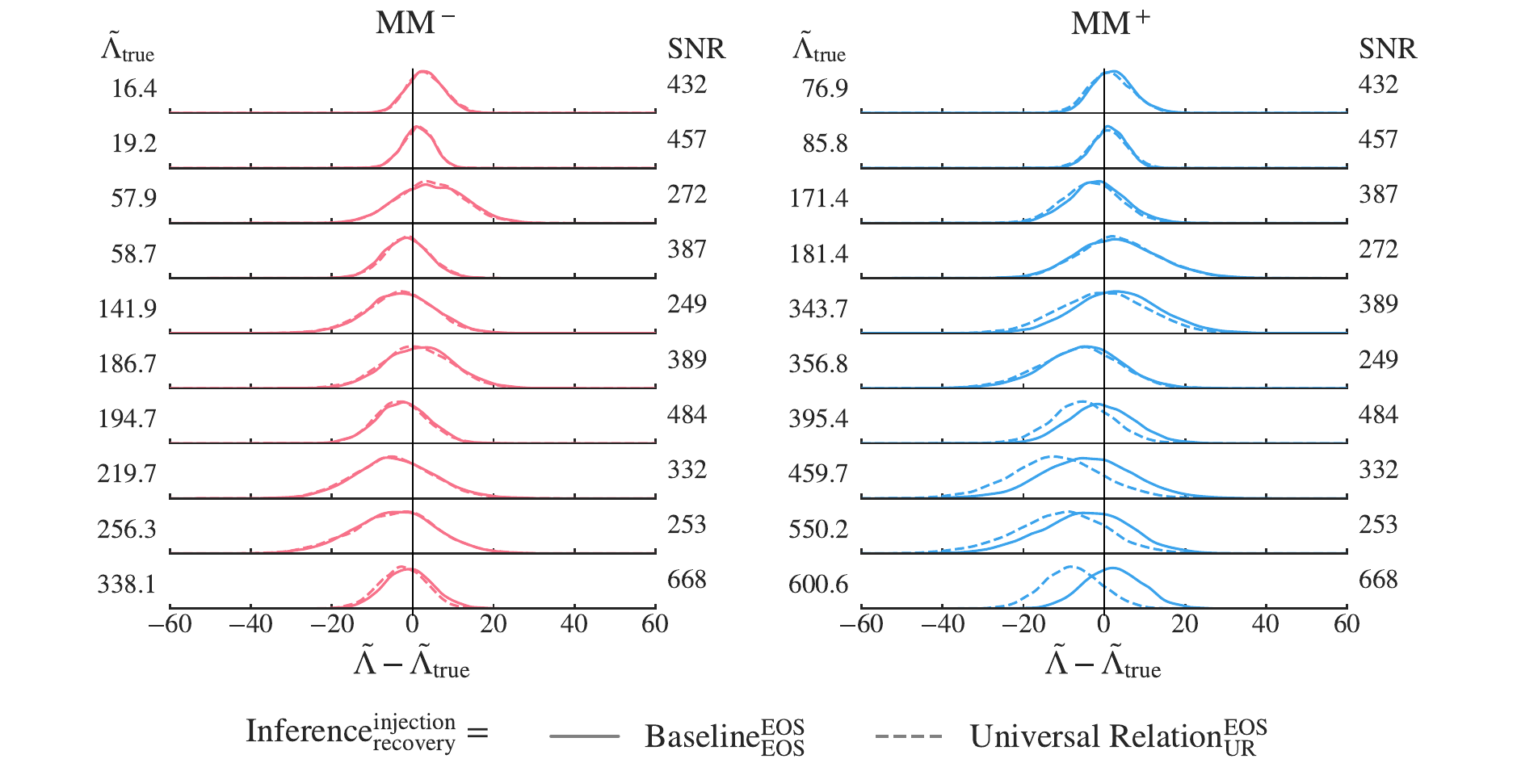}
\captionsetup{justification=raggedright,singlelinecheck=false}
    \caption{\textit{Fundamental mode frequency:} Posterior probability distributions of $\tilde{\Lambda} - \tilde{\Lambda}_{\mathrm{true}}$, where $\tilde{\Lambda}_{\mathrm{true}}$ is the true injected joint tidal deformability, for MM$^{-}$ (left) and MM$^{+}$ (right) EOS. An unbiased measurement corresponds to zero (black line). Results are shown for the ten highest-SNR events, ordered by the magnitude of the injected $\tilde{\Lambda}_{\mathrm{true}}$, with SNRs on the right of each posterior. Two analyses are compared as described in the text: Baseline (solid) and Universal Relation (dashed).}
    \label{fig:fpop}
\end{figure*}

\subsection{Population Draws}
Posteriors of $\tilde{\Lambda}$ relative to the injected value are shown for the ten loudest events in Fig.~\ref{fig:fpop}. As in the selected cases analyses, for MM$^-$ negligible bias is observed across all events, with consistent recovery of the injected $\tilde{\Lambda}$. In the case of MM$^+$ moderate biases towards a softer EOS are observed as $\tilde{\Lambda}$ increases beyond $>400$, though the injected value is recovered across all events to 90\% confidence. We note that the population here includes spin, however the dynamical tides model does not account for spin-dependent shifts in the $f$-mode frequency, so biases may be altered if these effects were to be included. While not shown in Fig.~\ref{fig:fpop}, as in the selected cases, the biases from using the original NRTidalv3 fit in the analysis exceed those from the qUR for all events, and in most cases the injected $\tilde{\Lambda}$ is not recovered.

\section{Binary Love}
\label{sec:BL}
\subsection{Implementation}
While the first two qURs concern the properties of isolated NSs, the Binary Love qUR focuses specifically on binary systems. This relation connects the symmetric $\Lambda_s \equiv (\Lambda_1+\Lambda_2)/2$ and the anti-symmetric $\Lambda_a \equiv (\Lambda_2-\Lambda_1)/2$ combinations of the components' tidal deformabilities and the mass ratio
\begin{equation}
\begin{split}
\Lambda_a(\Lambda_s,q;\vec{b})=&\frac{1-q^{10/(3-n)}}{1+q^{10/(3-n)}}\Lambda_s \times \\ 
&\frac{1+\sum_{i=1}^{3}\sum^{2}_{j=1}c_{ij}q^j\Lambda_s^{-i/5}}{1+\sum_{i=1}^{3}\sum^{2}_{j=1}d_{ij}q^j\Lambda_s^{-i/5}}\, ,
\end{split}
\end{equation}
where the parameters $\vec{b}=\{c_{ij}, d_{ij}\}$ are fitted to a wide range of EOSs and can be found in Ref.~\cite{Yagi:2015pkc}.

In contrast to the two previously discussed qURs, the LIGO Scientific, Virgo-, and Kragra Collaborations do not employ the binary Love relation by default when generating waveform or analysing data. However, the relation has been used when analysing real event data~\cite{Chatziioannou:2018vzf, Kumar:2019xgp, Chatterjee:2021xrm}, and is implemented within \textsc{Bilby}. The relation enables not sampling over both tidal deformabilities, but only $\Lambda_s$, with $\Lambda_a$ being inferred from the binary Love relation. This reduces the parameter space dimensionality, and implicitly assumes that all NSs share the same EOS. The binary Love qUR carries a larger intrinsic error due to EOS variation than the other qURs considered here, up to $\mathcal{O}(10\%)$. Therefore this error is accounted for by adding a Gaussian distribution within the calculation to mimic an average error $\sim10\%$
\begin{equation}
    \Lambda_a = \Lambda_a(\Lambda_s,q;\vec{b}) + \mathcal{N}(\mu(\Lambda_s,q), \sigma(\Lambda_s,q)) \, ,
\end{equation}
where the functions for $\mu, \sigma$ can be found in Ref.~\cite{Chatziioannou:2018vzf}.

\begin{figure}[htbp!]
    \centering
        \includegraphics[width=\columnwidth]{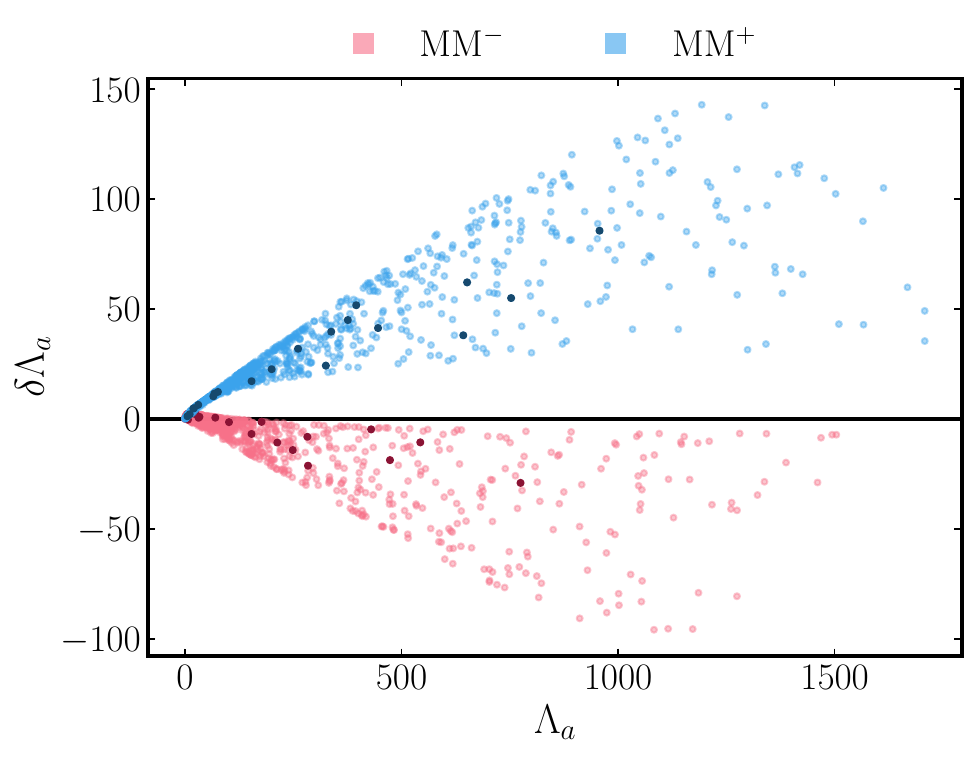}
    \captionsetup{justification=raggedright,singlelinecheck=false}
    \caption{Deviation $\delta\Lambda_a \equiv \Lambda_a^{\rm qUR}-\Lambda_a^{\rm EOS}$ of the anti-symmetric combination of tidal deformabilities $\Lambda_a$ from the predicted value from the binary Love qUR for MM$^{-}$ (pink) and MM$^{+}$ (blue) between masses of $1M_\odot$ and $2.2M_\odot$. Highlighted points show the 20 loudest sources of the population.}
    \label{fig:LaUR}
\end{figure}

Figure~\ref{fig:LaUR} shows the deviation between the value of $\Lambda_a$ predicted by the binary Love qUR and that computed directly from the EOS. The size of the deviation strongly depends on the mass ratio of the binary, leading to a wide spread across systems. For MM$^{-}$, $\Lambda_a$ is systematically underestimated, while for MM$^{+}$ it is overestimated, leading to relative errors in the component tidal deformabilities of up to 90\% and 64\% respectively in our population.

\subsection{Selected Cases}
\begin{figure}[htbp!]
        \centering
        \includegraphics[width=\columnwidth]{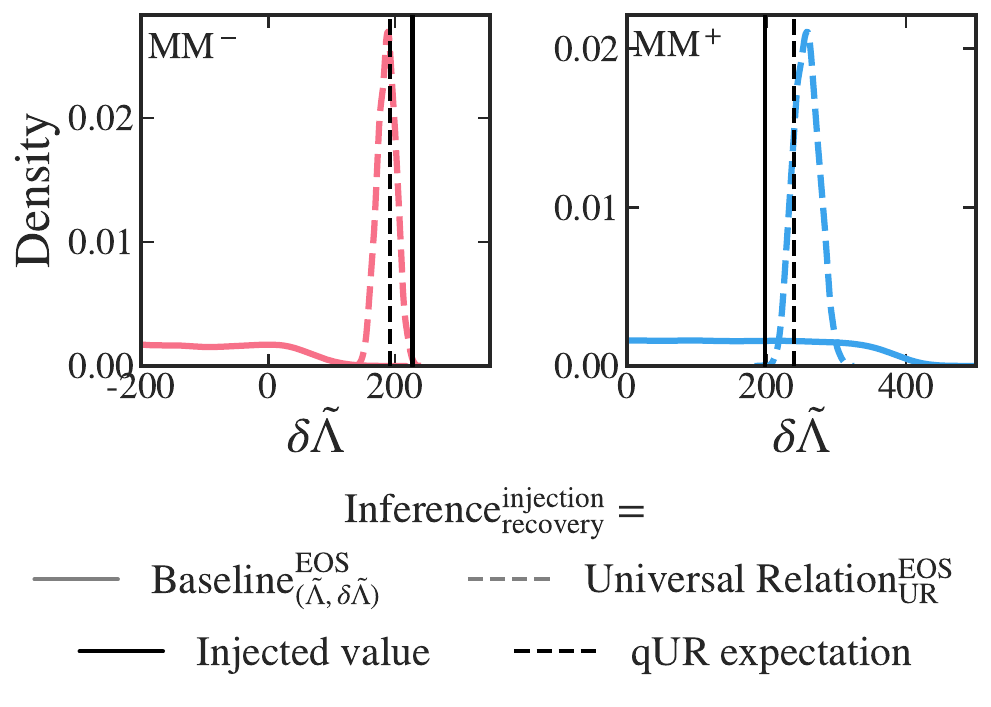}
        \captionsetup{justification=raggedright,singlelinecheck=false}
        \caption{\textit{Binary Love:} Posterior density distributions of $\delta\tilde{\Lambda}$ for the selected binary Love case with EOSs MM$^{-}$ (left) and MM$^{+}$ (right) for two assumptions: the signal is generated using $\Lambda_1$, $\Lambda_2$ directly computed from the EOS, with the analysis sampling tidal parameters independently (solid); the signal is generated using the EOS, but the analysis assumes the qUR (dashed). True injected values are indicated (black solid lines) alongside the value of $\delta\tilde{\Lambda}$ that would be obtained if calculated with the qUR is also shown (black dashed line).}
        \label{fig:Lapost}

\end{figure}

The non-spinning configurations that maximise the deviation from the qUR are found to be at $(m_1, m_2)=(1.29M_\odot, 1.03M_\odot)$ for MM$^{-}$ and $(m_1, m_2)=(1.68M_\odot, 1.04_\odot)$ for MM$^{+}$; cf.~Appendix~\ref{Appdx:mismatches}.

We perform two sets of analyses for each of these configurations:
\begin{itemize}
\item \textit{Baseline}: We simulate the waveform using the tidal deformabilites obtained from the EOS, and perform parameter estimation sampling independently in $(\tilde{\Lambda}, \delta\tilde{\Lambda})$. This serves as a reference point for comparison.
\item \textit{Universal Relation}: We simulate the waveform using the tidal deformabilites obtained from the EOS, and perform parameter estimation such that only $\Lambda_s$ is sampled over, $\Lambda_a$ is computed using the qUR. This allows us to quantify biases arising from incorrectly assuming the qUR.
\end{itemize}

We find that the leading-order tidal parameter $\tilde{\Lambda}$ is recovered without significant bias for both EOSs. However, the next-to-leading-order parameter $\delta\tilde{\Lambda}$ exhibits systematic biases when the binary Love qUR is applied, as shown in Fig.~\ref{fig:Lapost}. With independent sampling, $\delta\tilde{\Lambda}$ is very poorly constrained even for third-generation detectors: the posterior for MM$^+$ is broad, but contains the true value to 90\% confidence, while for MM$^-$ the true value is not recovered. By contrast, applying the qUR drastically narrows the posteriors, but the recovered $\delta\tilde{\Lambda}$ is biased toward the qUR prediction. This demonstrates a trade-off inherent to qURs, namely, the reduction of the dimensionality of the parameter space decreasing the measurement uncertainty, while introducing potential systematic biases. While $\tilde{\Lambda}$ remains unbiased, the bias in $\delta\tilde{\Lambda}$ propagates to the individual tidal deformabilities $\Lambda_1$ and $\Lambda_2$, which are necessary for accurate EOS constraints. These results highlight that caution is needed when relying on the binary Love relation for next-to-leading-order tidal measurements and related inferences.

\subsection{Population Draws}
\begin{figure*}
    \centering
    \includegraphics[width=0.9\linewidth]{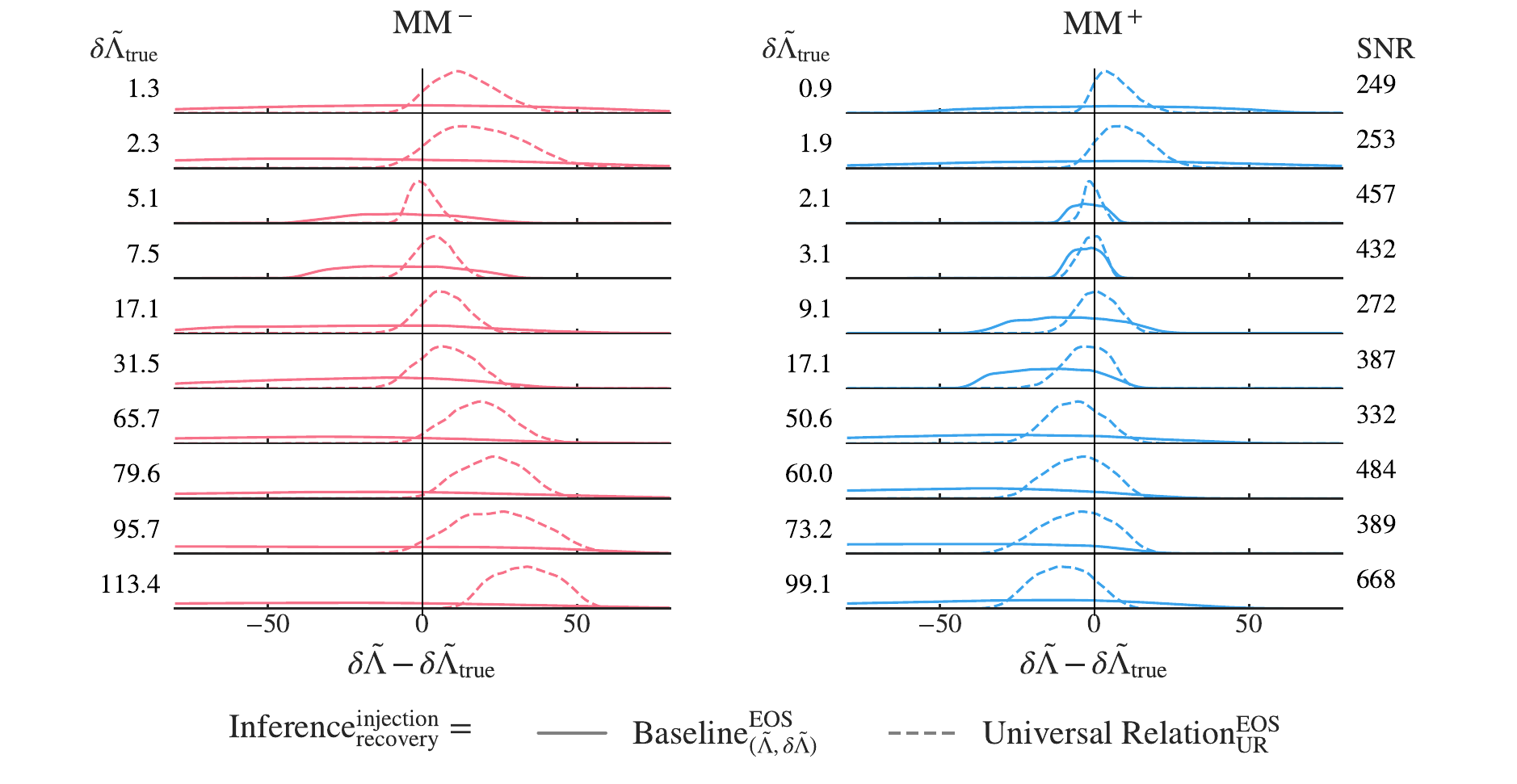}
\captionsetup{justification=raggedright,singlelinecheck=false}
    \caption{\textit{Binary Love:} Posterior distributions of $\delta\tilde{\Lambda} - \delta\tilde{\Lambda}_{\mathrm{true}}$ for MM$^{-}$ (left) and MM$^{+}$ (right), with $\delta\tilde{\Lambda}_{\mathrm{true}}$ given by the EOS. The vertical black line at zero denotes unbiased recovery. Results are presented for the ten highest-SNR events, ordered according to the magnitude of the injected $\delta\tilde{\Lambda}_{\mathrm{true}}$, with the SNRs of each row shown on the right hand side. Two analyses are compared as described in the text: Baseline (solid) and Universal Relation (dashed)}
    \label{fig:Lapop}
\end{figure*}

Posterior distributions of $\delta\tilde{\Lambda}$ for the ten loudest zero-noise signals in the population are shown in Fig.~\ref{fig:Lapop}. Signals injected with tidal deformabilities derived from the EOS show broad posteriors when tidal deformabilities are sampled independently, although consistently here contain the true values. Assuming the binary Love qUR in the recovery introduces mild systematic biases that grow with $\delta\tilde{\Lambda}$, tending positive for MM$^{-}$ and negative for MM$^{+}$ as expected from the direction of the qUR deviation from Fig~\ref{fig:LaUR}.

\section{Hierarchical EOS Inference}
\label{sec:pop}
Until now, we have focused on the potential biases in the recovery of individual source parameters. However, as the number of observed BNS mergers increases with third-generation sensitivity, we can move beyond single-event analyses to infer the underlying EOS through a hierarchical combination of multiple detections. Hence, we now combine the 20 individual-event posteriors to obtain a population-level constraint on the EOS, examining how the assumed qURs affect the joint result. Previously, we considered zero-noise simulations to isolate the effect of the qURs. To obtain more realistic population-level EOS constraints, we now include simulated detector noise, with each event assigned a random noise realization held fixed across all tested qURs. 

We perform four sets of analyses on the population:
\begin{itemize}
\item \textit{Baseline}: The reference analysis uses the default implementation of \textsc{NRTidalv3} for both injection and recovery. The SIQM and the $f$-mode frequencies entering the dynamical tides are both obtained from their respective qUR, and tidal deformabilities are sampled independently in $(\tilde{\Lambda},\delta\tilde{\Lambda})$.
\item \textit{Quadrupole}: The injection waveform uses SIQM values obtained from the EOS fit, while the analysis waveform is identical to the baseline, employing the default \textsc{NRTidalv3} model and assuming the SIQM qUR.
\item \textit{Fundamental mode}: Both injection and recovery replace the default \textsc{NRTidalv3} dynamical tides with our refitted dynamical tide models. The injection waveform uses EOS-based $f$-mode frequencies, while the recovery assumes $f$-mode frequencies from the qUR.
\item \textit{Binary Love}: The injection waveform is identical to the baseline with independent tidal deformabilities $(\Lambda_1,\Lambda_2)$. The recovery analysis instead enforces the binary Love qUR by sampling tides only the symmetric tidal parameter $\Lambda_s$.
\end{itemize}

We follow the hierarchical population methodology outlined in Sec.~\ref{subsec:population}. Given the spectral coefficients we sample in, we solve the Tolman-Oppenheimer-Volkoff~\cite{Oppenheimer:1939ne} equations as implemented in \textsc{LALSuite} to map the EOS samples to $m-\Lambda$ samples. 

\begin{figure*}
    \centering
    \includegraphics[width=\linewidth]{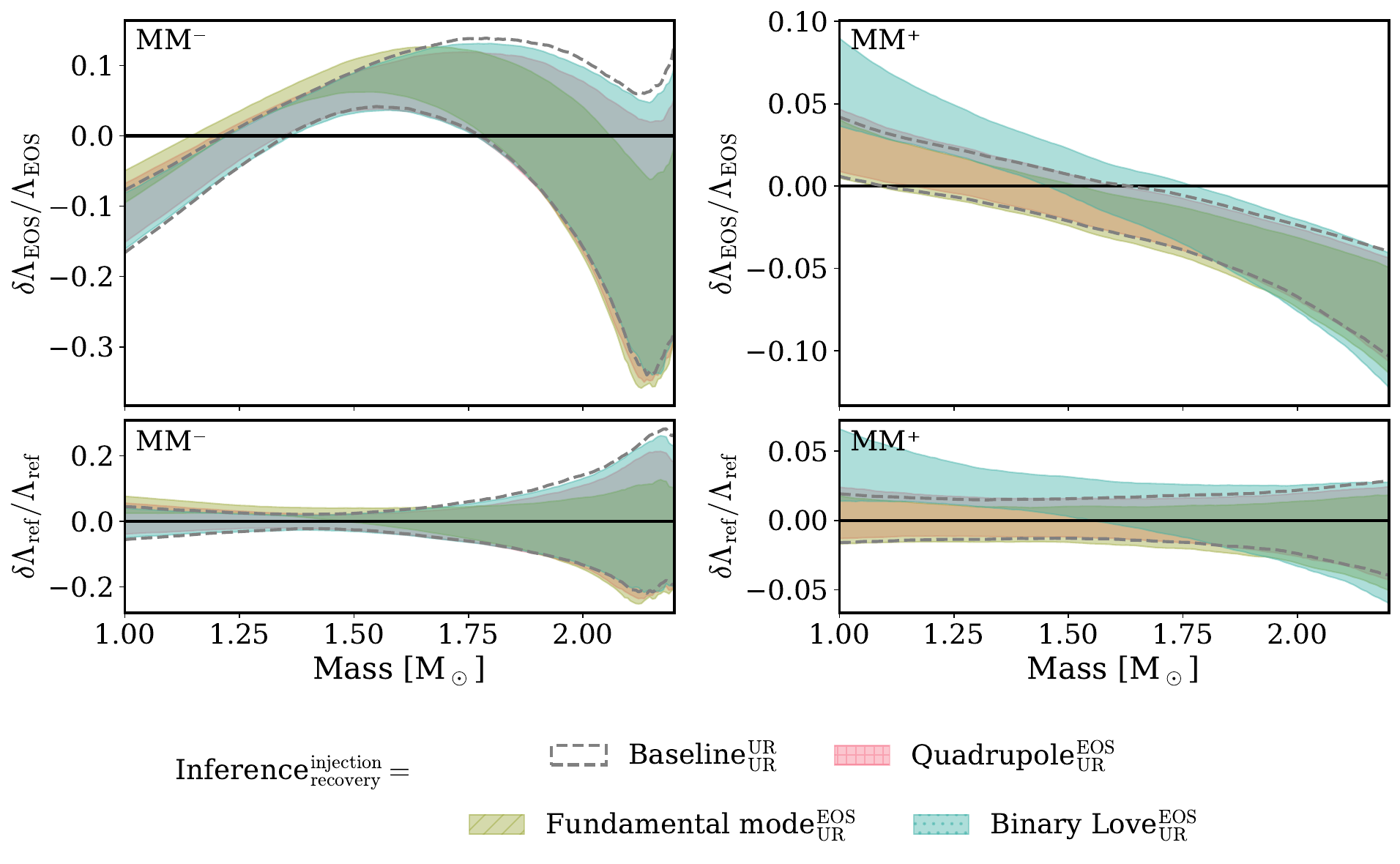}
\captionsetup{justification=raggedright,singlelinecheck=false}
    \caption{EOS constraints from the population of the 20 highest SNR events, showing 90\% confidence intervals on the mass–$\Lambda$ relation for MM$^{-}$ (left) and MM$^{+}$ (right). Coloured regions correspond to recoveries assuming different qURs: quadrupole (pink), $f$-mode frequency (green), and binary Love (blue). As a reference, the outcome of baseline runs are shown where both injection and recovery assumed the quadrupole and $f$-mode qURs with independent tidal parameters (grey dashed line). \textit{Top panel:} The fractional difference between the recovered $\Lambda$ and the true injected EOS $\Lambda_{\rm EOS}$, $\delta\Lambda_{\rm EOS}/\Lambda_{\rm EOS}\equiv(\Lambda-\Lambda_{\rm EOS})/\Lambda_{\rm EOS}$. True agreement with the EOS is shown alongside (black solid line). \textit{Bottom panel:} The fractional difference between the recovered $\Lambda$ and the median of the baseline posterior $\Lambda_{\rm ref}$, $\delta\Lambda_{\rm ref}/\Lambda_{\rm ref}\equiv(\Lambda-\Lambda_{\rm ref})/\Lambda_{\rm ref}$.  True agreement with the median of the baseline is shown alongside (black solid line)} 
    \label{fig:EOSposterior}
\end{figure*}

We show the 90\% confidence intervals for the reconstructed $m-\Lambda$ samples in Fig.~\ref{fig:EOSposterior}, relative to the true injected EOSs. In the top panels, we show the fractional difference between the recovered posteriors and the injected EOS. For the baseline analysis, the true EOS lies outside the confidence bounds across much of the mass range. We attribute this discrepancy to differences in the spectral EOS parametrisation used in the analysis compared with the MM approach used for the simulated signals. For reference, we also show the posteriors difference relative to the median of the baseline to illustrate the relative impact of the qUR assumptions on the reconstructed EOS.

We compare the resulting posterior 90\% confidence intervals obtained when assuming each qUR investigated in this study. When considering the population with the MM$^-$ EOS, we observe that assuming either the SIQM or binary Love qURs have negligible impact on the EOS $m-\Lambda$ constraints relative to the baseline analysis. We observe a minor bias on the posterior obtained with the fit assuming the $f$-mode qUR, for which the tidal deformability is slightly overestimated for masses $\lesssim 1.75 M_{\odot}$, however this result is still consistent to 90\% confidence with the baseline posterior. For the MM$^+$ EOS, assuming the SIQM or $f$-mode frequency qUR has a negligible effect on the posterior. Assuming the binary Love qUR for the MM$^+$ EOS results in an overestimation $\sim$5\% at 1$M_{\odot}$, which steadily decreases so that for masses $\gtrsim 1.75 M_{\odot}$ the posterior is broadly consistent with that of the baseline. This demonstrates recovery of a stiffer EOS for lower masses.

\section{Conclusions}
\label{sec:conclusions}
The use of qURs in BNS waveform data analysis is a valuble tool for reducing degeneracies and the parameter space dimensionality without sacrificing waveform accuracy in current detectors. Here, we assess the use of qURs in the context of the third-generation era to determine whether they remain robust when accounting for extreme but observationally consistent EOSs. We investigate three qURs, and for each, we conduct a parameter estimation campaign in which we consider a) selected cases for systems devised such that the deviation from qURs is largest, b) population draws from an astrophysically motivated population, and c) hierarchical population inference. We find that for each qUR considered:

    \textit{Spin-induced quadrupole moment}: For low-spin systems typical of the BNS population, the SIQM qUR introduces minimal biases. However, at even moderate spins of $|\chi|=0.15$ we find appreciable biases across multiple intrinsic parameters, including the masses, spins, and tidal deformabilities. Allowing $C_Q$ to vary independently largely mitigates these biases. These results are consistent with the known limitations of the SIQM qUR, which breaks down outside of the slow-rotation regime~\cite{Doneva:2013rha}, and suggest that future waveform models for spinning NSs should incorporate rotation-dependent extensions of the qUR~\cite{Pappas:2013naa}.

    \textit{Fundamental mode frequency}: For individual events, the $f$-mode qUR can produce minor shifts in recovered tidal deformability, biasing them toward softer values for one EOS and producing negligible shifts for the other. At the population level, however, posteriors remain consistent with the baseline, indicating that overall the qUR introduces a negligible bias. By comparison, waveform systematics associated with the phenomenological frequency-domain dynamical tide fit in \textsc{NRTidalv3} dominate, and push tidal deformabilities toward a stiffer EOS. Future work could explore two complementary directions: first, developing improved or recalibrated $f$-mode qURs to further reduce the already modest qUR-induced biases~\cite{Raithel:2022orm}; and second, waveform models that use the $f$-mode frequency directly in each waveform evaluation~\cite{Williams:2022vct, Schmidt:2019wrl, Pratten:2019sed}, avoiding an unnecessary phenomenological fit and reducing waveform systematics associated with dynamical tides.

    \textit{Binary Love}: The binary Love qUR enables measurement of next-to-leading-order $\delta\tilde{\Lambda}$ that are otherwise largely unconstrained, but introduces systematic biases that propagate to the individual tidal deformabilities. For the population studied here, the bias is minor to moderate, with one EOS demonstrating a bias at lower masses. Using the qUR does not lead to increased accuracy in EOS measurements, indicating that the primary benefit of the relation is in individual-event inference. These results highlight that the binary Love qUR is a valuable tool for extracting tidal information, but its limitations should be considered for analyses that rely on the components $\Lambda$, for example when interpreting extreme EOSs.

Our findings demonstrate that while qURs offer a convenient means to reduce dimensionality, they may introduce modest biases in certain cases and thus warrant careful attention. At the population level, however, these biases are generally negligible, and more significant errors arise from waveform systematics, which should be prioritised in future improvements. Future extensions to this work could explore universal relations beyond those considered here, such as higher multipole tidal and gravito-magnetic effects, as well as improved or recalibrated qURs~\cite{Godzieba:2020bbz, Raithel:2022orm, Doneva:2013rha, Kruger:2025sda, Papigkiotis:2025dka} to reduce biases. Investigating the role of qURs within alternative tidal waveform models~\cite{Haberland:2025luz,Gamba:2023mww, Williams:2024twp}, considering precessing spin configurations or higher order modes~\cite{Abac:2025brd}, and extending to EOSs with phase transitions~\cite{Bauswein:2018bma} are all promising directions to better understand the limitations and applicability of qURs for future detections.

\section*{Acknowledgments}
The authors thank Hauke Koehn and Henrik Rose for useful discussions and comments. We thank Adrian Abac for providing fitting routines for \textsc{NRTidalv3} dynamical tides phenomenological fit. We also thank Patricia Schmidt and Geraint Pratten for useful discussions in an earlier project related to this work. The authors additionally thank Lami Suleiman for valued comments on the manuscript.

T.D. and A.P. acknowledge funding from the EU Horizon under ERC Starting Grant, no. SMArt-101076369. Views and opinions expressed are those of the authors only and do not necessarily reflect those of the European Union or the European Research Council. Neither the European Union nor the granting authority can be held responsible for them. 

Computations were performed on the DFG-funded research cluster Jarvis at the University of Potsdam (INST 336/173-1; project number: 502227537).

This research has made use of data or software obtained from the Gravitational Wave Open Science Center (gwosc.org), a service of the LIGO Scientific Collaboration, the Virgo Collaboration, and KAGRA. This material is based upon work supported by NSF's LIGO Laboratory which is a major facility fully funded by the National Science Foundation, as well as the Science and Technology Facilities Council (STFC) of the United Kingdom, the Max-Planck-Society (MPS), and the State of Niedersachsen/Germany for support of the construction of Advanced LIGO and construction and operation of the GEO600 detector. Additional support for Advanced LIGO was provided by the Australian Research Council. Virgo is funded, through the European Gravitational Observatory (EGO), by the French Centre National de Recherche Scientifique (CNRS), the Italian Istituto Nazionale di Fisica Nucleare (INFN) and the Dutch Nikhef, with contributions by institutions from Belgium, Germany, Greece, Hungary, Ireland, Japan, Monaco, Poland, Portugal, Spain. KAGRA is supported by Ministry of Education, Culture, Sports, Science and Technology (MEXT), Japan Society for the Promotion of Science (JSPS) in Japan; National Research Foundation (NRF) and Ministry of Science and ICT (MSIT) in Korea; Academia Sinica (AS) and National Science and Technology Council (NSTC) in Taiwan.

\appendix

\section{Equation of State Construction}
\label{Appdx:mm}

For the construction of the EOSs employed in this work, we start by expressing the energy per particle as a Taylor expansion around the saturation density $n_\mathrm{sat}$,
\begin{eqnarray}
e_\mathrm{SNM}(n) &=& E_\mathrm{sat} + \frac{1}{2} K_\mathrm{sat} x^2 + \frac{1}{6} Q_\mathrm{sat} x^3 + \frac{1}{24} Z_\mathrm{sat} x^4 + \dots , \nonumber \\
e_\mathrm{sym}(n) &=& E_\mathrm{sym} + L_\mathrm{sym} x + \frac{1}{2} K_\mathrm{sym} x^2 + \frac{1}{6} Q_\mathrm{sym} x^3 \nonumber \\
&&+ \frac{1}{24} Z_\mathrm{sym} x^4 + \dots ,
\end{eqnarray}
where $e_\mathrm{SNM}$ and $e_\mathrm{sym}$ denote the energy per particle of symmetric nuclear matter and the symmetry energy, respectively.
The density expansion parameter is $x = (n - n_\mathrm{sat}) / (3 n_\mathrm{sat})$, with $n=n_p+n_n$ the baryon density.
The coefficients $E_\mathrm{sat}$, $ K_\mathrm{sat}$, $ Q_\mathrm{sat}$, $ Z_\mathrm{sat}$, $ E_\mathrm{sym}$, $ L_\mathrm{sym}$, $ K_\mathrm{sym}$, $ Q_\mathrm{sym}$, $ Z_\mathrm{sym}$ are the nuclear empirical parameters~\cite{Margueron2018a}.
The total energy per nucleon of asymmetric (infinity) nuclear matter then reads, 
\begin{equation}
e(n,\delta) = e_{\text{SNM}}(n) + \delta^2 e_{\text{sym}}(n).
\label{eq:emm}
\end{equation}
where $\delta=(n_n - n_p)/n$ is the nuclear asymmetry. This expansion inspired the MM, in which the model energy per particle is $e_\mathrm{MM} = e_\mathrm{kin} + e_\mathrm{pot}$, with the kinetic energy given by a Fermi gas of non-relativistic nucleons. The potential energy is a series expansion around saturation. Therefore, the model parameters are the empirical parameters expressed above. We write the model parameters of the two EOS used in the present work in Table~\ref{tab:NEP}. More details of the MM are in Ref.~\cite{Margueron2018a}.

All EOSs assume cold, catalyzed, charge-neutral matter composed of neutrons, protons, electrons, and muons in $\beta$-equilibrium.
The equilibrium conditions are given by
\begin{equation}
\mu_n = \mu_p + \mu_e, \qquad \mu_e = \mu_\mu,
\label{eq:betaeq}
\end{equation}
where $\mu_i$ denotes the chemical potential of species $i = n, p, e, \mu$.
Electrons and muons are treated as relativistic fermion gases as in Refs.~\cite{Grams22a, Grams22b}, and neutrinos are neglected since they do not participate in equilibrium for cold NSs.
Equations~\eqref{eq:betaeq}, together with charge neutrality ($n_e + n_\mu = n_p$) and baryon number conservation ($n_B = n_n + n_p$), determine the matter composition at each baryon density.

Because the MM is based on a non-relativistic kinetic term, it may violate causality at high densities.
To prevent this, we adopt a constant-sound-speed extension beyond the central density corresponding to a $2.2~M_\odot$ NS.
Above this threshold, the sound speed $c_s$ is kept constant, and the EOS is obtained by integrating
\begin{equation}
 \left(  \frac{c_s}{c} \right)^2 =\frac{d P}{d\rho} \;,
\end{equation}
where $P$ is the pressure, $\rho$ and energy density, and  $c$ is the speed-of-light, which is set to one.
The resulting EOS is then used to solve the Tolman–Oppenheimer–Volkoff (TOV)~\cite{Tolman1939,Oppenheimer1939}, tidal deformability, and $f$-mode oscillation equations.

In summary, both EOSs constructed for the present work use the MM up to the central density of a $2.2~M_\odot$~NS, above which the construction transitions to the constant–sound-speed extension.

\begin{table}[t]
\centering
\captionsetup{justification=raggedright,singlelinecheck=false}
\caption{\label{tab:NEP}
Parameters for the two metamodel EoS used in the present work. }
\begin{tabular}{c|ccl}
\hline\hline
Parameter  & MM$^{-}$ & MM$^{+}$ \\
\hline
$E_\sat$  [MeV]       & -16.024 & -16.073 \\
$n_\sat$  [fm$^{-3}$] & 0.1491 & 0.1468 \\
$K_\sat$  [MeV]      & 249.4 & 226.8 \\
$Q_\sat$  [MeV]      & -218.1 & 778.8 \\
$Z_\sat$  [MeV]      & -817.6 & 234.1 \\
$E_\sym$  [MeV]      & 36.54 & 24.68 \\
$L_\sym$  [MeV]      & 76.38 & 26.70 \\
$K_\sym$  [MeV]      & -169.56 & -108.07 \\
$Q_\sym$  [MeV]      & -159.1 & 8.543 \\
$Z_\sym$  [MeV]      & 860.5 & 955.5 \\
\hline\hline
\end{tabular}
\end{table}


\section{Dynamical Tides Refitting in \textsc{NRTidalv3}}
\label{Appdx:dynamicaltides}

Dynamical tides are incorporated into \textsc{NRTidalv3} through a time-dependent enhancement factor to the quadrupole Love number, $k_2$. This enhancement transforms as $k_2 = k_2(\omega)\rightarrow k_2k_2^{\rm eff}(\omega)$ where $\omega = d\phi/dt$ denotes the instantaneous GW frequency derived from the time-domain phase $\phi$. The enhancement factor is derived from a dynamical quadrupole moment obeying the equation of motion of a tidally driven harmonic oscillator with relativistic corrections, and is inherently a time-dependent quantity. However, it is more efficient to construct waveform models in the frequency domain, as the frequency dependent power spectral density must be incorporated into each likelihood evaluation. The frequency-domain analytical PN expression of the tidal phase can be written as
\begin{equation}
    \psi^{\rm PN}_T = \bar{k}^{\rm eff}_{2}(\hat{\omega})\psi^{\rm PN}_T(\hat{\omega}),
    \label{eq:pn-phase}
\end{equation}
where $\hat{\omega}$ is a rescaled dimensionless gravitational-wave frequency $\hat{\omega}=M\omega$ with $M$ being the total, and $\bar{k}^{\rm eff}_{2}$ is the frequency-domain enhancement factor. The frequency-domain phase $\psi$ is obtained from the time-domain one through the stationary-phase approximation
\begin{equation}
\frac{d^2\psi(\omega)}{d\omega^2}    = \frac{1}{\omega}\frac{d^2\phi(\omega)}{d\omega^2}.   
\end{equation}
Using the PN approximation in Eq.~\ref{eq:pn-phase} yields a second-order differential equation
\begin{equation}
    \frac{d^2 \bar{k}^{\rm eff}_{2}}{d\hat{\omega}^2}\psi_T^{\rm PN} + 2  \frac{d \bar{k}^{\rm eff}_{2}}{d\hat{\omega}}\frac{d \psi_T^{\rm PN}}{d\hat{\omega}} + \bar{k}^{\rm eff}_{2}\frac{d^2 \psi_T^{\rm PN}}{d\hat{\omega}^2} = \frac{1}{\hat{\omega}}\frac{d \phi_T^{\rm PN}}{d\hat{\omega}}\, .
\end{equation}
from which the frequency-domain enhancement factor can be computed imposing boundary conditions
\begin{equation}
    \bar{k}_2^{\rm eff}(0)=1, \qquad \frac{d\bar{k}_2^{\rm eff}(0)}{d\hat{\omega}}=0,
\end{equation}
yields $\bar{k}^{\rm eff}_2$. However, numerically solving this per waveform evaluation is computationally inefficient and not practically feasible. Therefore a phenomenological representation is obtained through fitting $\bar{k}^{\rm eff}_2$ as a function of the rescaled gravitational wave frequency. This takes the form
\begin{equation}
\begin{split}
\bar{k}_{2, \rm rep}^{\rm eff} =& 1 - \frac{s_1-1}{\exp[-s_2(\hat{\omega} - s_3)]+1} 
- \frac{s_1-1}{\exp[s_2 s_3]+1} \\
& - \frac{s_2 (s_1-1) \exp[s_2 s_3]}{(\exp[s_2 s_3]+1)^2} \hat{\omega},
\label{eq:phenomfit}
\end{split}
\end{equation}
where parameters $s_i$,  $(i=1,2,3)$ are constrained using $s_{i,j}$, $(j=1,2,3)$:
\begin{equation}
    s_i = s_{i,0}+s_{i,1}\kappa_{\rm eff} + s_{i,2}q\kappa_{\rm eff}
    \label{eq:phenomparams}
\end{equation}
and $\kappa_{\rm eff}$ is the effective tidal parameter
\begin{equation}
    \kappa_{\rm eff} = \frac{2}{13}\bigg\{ \bigg[1 + \frac{12m_B}{m_A}\bigg(\frac{m_A}{C_A(m_1+m_2)}k_2^A\bigg)\bigg]+[A\leftrightarrow B \,]\bigg\}\, .
\end{equation}

The original \textsc{NRTidalv3} fit for $\bar{k}_2^{\rm eff}$ is calibrated over 55 numerical-relativity waveforms, spanning a wide range of EOSs and NS masses, using $f$-mode frequencies from the qUR. To isolate the effect of the $f$-mode qUR, we follow the procedure laid out in \cite{Abac:2023ujg} to recalibrate the phenomenological model\footnote{Note that we recalibrate only the dynamical tide prefactor, and we do not recalibrate the full \textsc{NRTidalv3} phase.} \eqref{eq:phenomfit} twice for each EOS on a uniform mass grid $m\in[1.0,2.2]\ M_\odot$: once using EOS-based $f$-mode frequencies and once again using frequencies from the qUR, in order to isolate the effects of the qUR. We write the parameters for these analytical fits in Tab.~\ref{tab:phenomparams}. These fits reproduce the numerically calculated $\bar{k}_2^{\rm eff}$ to within $\sim$2\%, comparable to the reported accuracy of the original \textsc{NRTidalv3} fit on its calibration dataset. 

\begin{figure*}
    \centering
    \includegraphics[width=\linewidth]{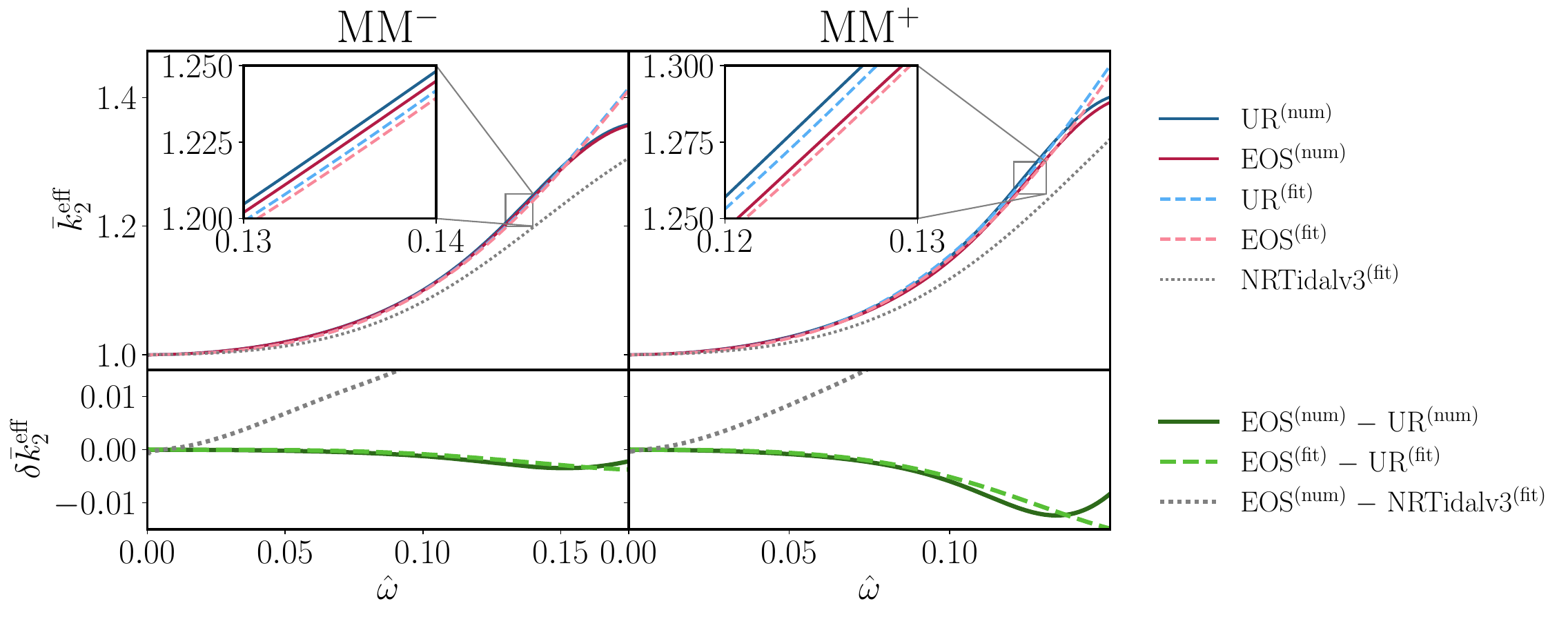}
\captionsetup{justification=raggedright,singlelinecheck=false}
    \caption{\textit{Top}: Effective dynamical Love number $\bar{k}_{2}^{\rm eff}$ for an equal-mass $1.4M_\odot+1.4M_\odot$ binary, shown for MM$^{-}$ (left) and MM$^{+}$ (right). Solid curves show numerically computed results, while dashed curves denote the phenomenological representation $\bar{k}_{2,\mathrm{rep}}^{\rm eff}$. In each case, the $f$-mode frequency is either taken from the EOS (red) or inferred from the qUR (blue). The original \textsc{NRTidalv3} fit is shown for comparison (grey dotted). \textit{Bottom}: Residuals between EOS- and qUR-based $f$-mode implementations for the numerically computed $\bar{k}_{2}^{\rm eff}$ (solid green) and for the phenomenological fits $\bar{k}_{2,\mathrm{rep}}^{\rm eff}$ (dashed green). Also shown is the residual between the EOS-based numerical result and the original \textsc{NRTidalv3} fit (grey dotted).}
    \label{fig:k2fit}
\end{figure*}

Figure~\ref{fig:k2fit} shows the numerically solved $\bar{k}_2^{\rm eff}$ for each EOS, computed using $f$-mode frequencies from the EOS itself and from the UR, along with the corresponding phenomenological fits for both cases. The original \textsc{NRTidalv3} fit is also shown for comparison. The qUR systematically underestimates the $f$-mode frequency (see Fig.~\ref{fig:fUR}), which should enhance the dynamical tidal effect, and this is reflected in the small increase of $\bar{k}_2^{\rm eff}$ for the qUR-based calculations compared to the EOS-based ones. While the phenomenological fits do not exactly reproduce the numerical results, the EOS–qUR residuals are comparable between the fits and the numerical calculations, indicating that the relative effect of the qUR is preserved for bias assessment. By contrast, for the extreme EOSs examined in this study, the original \textsc{NRTidalv3} fit substantially underestimates the dynamical tidal enhancement, reflecting a waveform systematic arising from its calibration across a range of EOSs which does not include the EOSs considered here. This propagates through the the observed constraints, as discussed in the main text.

\begin{table*}[t]
\centering
\captionsetup{justification=raggedright,singlelinecheck=false}
\caption{
Fitting coefficients $s_{i,j}$ for Eq.~\ref{eq:phenomparams} used in the phenomenological representation of the effective dynamical Love number in Eq.~\ref{eq:phenomfit}. 
For each EOS (MM$^{-}$ and MM$^{+}$), coefficients are shown for refits using $f$-mode frequencies computed directly from the EOS and inferred from the qUR. }

\label{tab:phenomparams}

\begin{tabular}{c | c || c c c || c c c}
\hline\hline
& & \multicolumn{3}{c||}{MM$^{-}$} & \multicolumn{3}{c}{MM$^{+}$} \\
\hline
$f$-mode & $i\backslash j$ & 0 & 1 & 2 & 0 & 1 & 2 \\
\hline
\multirow{3}{*}{EOS}
& 0 & 1.22789431 & 0.00843780563 & 0.08070611
    & 1.26016532 & 0.00923850345 & 0.00307465091 \\
& 1 & 22.6411246 & 0.00266341177 & $-0.0229058152$
    & 24.5049251 & 0.00975118999 & $-0.0149744133$ \\
& 2 & 0.169432184 & $-0.000129817652$ & $0.000294531211$
    & 0.156750970 & $-0.0000514905274$ & $0.000129091585$ \\
\hline
\multirow{3}{*}{qUR}
& 0 & 1.22827517 & 0.00820111996 & 0.0108501483 & 1.26318668 & 0.00874584407 & 0.00331920278 \\
& 1 & 22.8304860 & 0.00124936985 & $-0.0209458222$ & 24.8354839 & 0.00885295938 & $-0.0109442546$ \\
& 2 & 0.168708968 & $-0.000118521922$ & 0.000273745415 & 0.155265748 & $-0.0000535049515$ & 0.000114110940 \\
\hline\hline
\end{tabular}

\end{table*}

\section{Mismatches across Quasi-Universal Relations}
\label{Appdx:mismatches}
Here we present waveform mismatches for each qUR studied in this work using \textsc{IMRPhenomXAS\_NRTidalv3}, following the procedure outlined in Sec.~\ref{subsec:extreme_cases_method}. Mismatches are evaluated over the frequency range $f\in[5, 2048]$\,Hz with a sampling rate of 4096\,Hz on a uniform 200$\times$200 grid of component masses $m \in [1.0, 2.2],M_\odot$ covering the region where both EOSs are valid. All parameters not directly involved in the qUR studied are held fixed, with spins set as specified for each case and extrinsic parameters fixed to GW170817-like values, ensuring that waveform differences reflect only the qUR deviations. The specific setups are as follows:
\begin{itemize}
    \item \textit{SIQM $(\chi_1=\chi_2=0.15)$:} Compare waveforms using the SIQM qUR to those using EOS-derived SIQM values obtained through the RNS-based procedure of Appendix~\ref{Appdx:SIQM}.
    \item \textit{$f$-mode frequency $(\chi_1=\chi_2=0)$:} Compare waveforms using the refitted dynamical tides from App.~\ref{Appdx:dynamicaltides}, with $f$-mode frequencies taken from either the EOS or the qUR.
    \item \textit{Binary Love $(\chi_1=\chi_2=0)$:} Compare waveforms where $\Lambda_s$ is taken from the EOS and $\Lambda_a$ is set via the binary Love qUR and mapped $(\Lambda_a,\Lambda_s)\rightarrow(\Lambda_1,\Lambda_2)$, to waveforms using $\Lambda_1, \Lambda_2$ directly set from the EOS.
\end{itemize}
\begin{figure*}
    \centering
    \includegraphics[width=\linewidth]{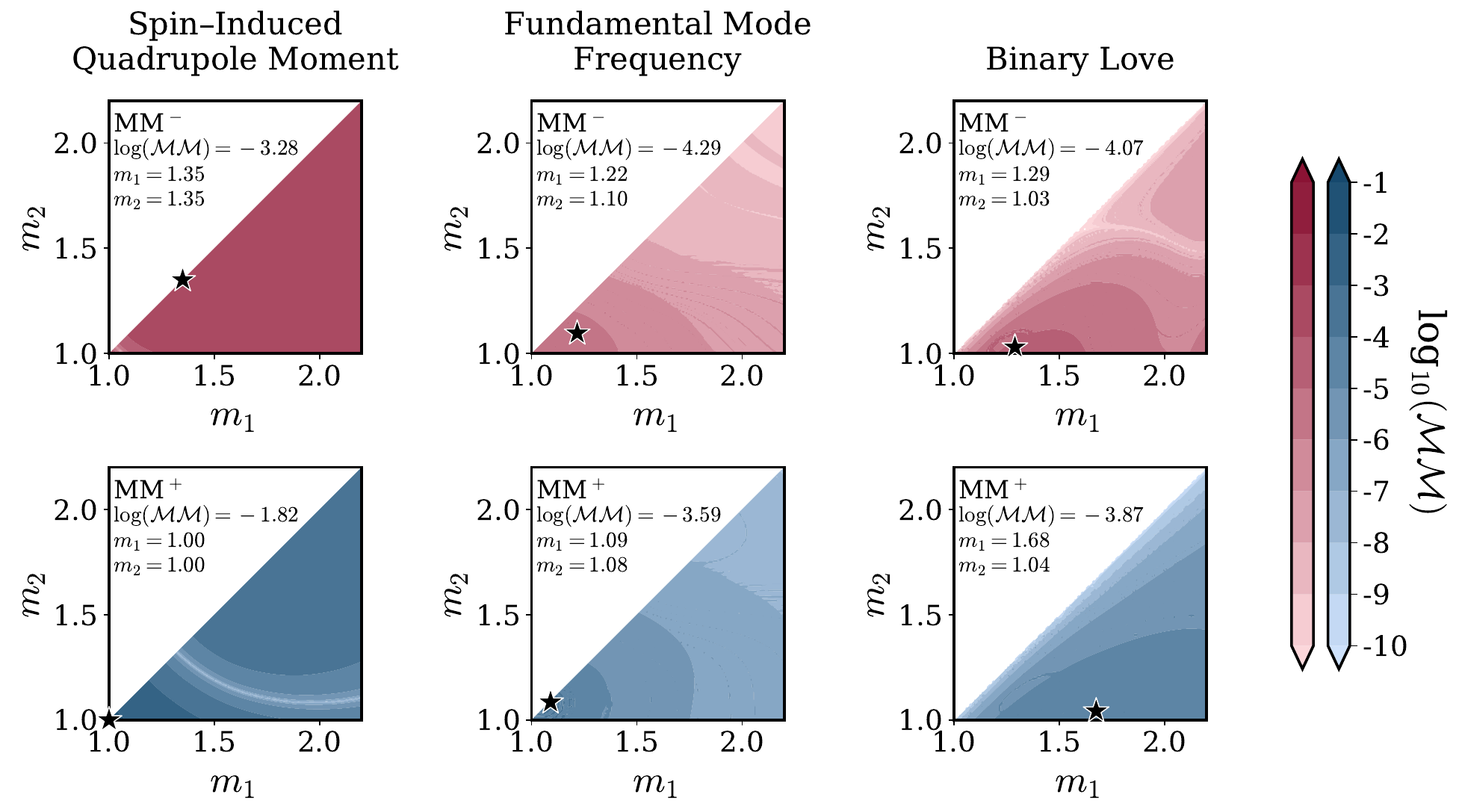}
\captionsetup{justification=raggedright,singlelinecheck=false}
    \caption{Mismatches between the waveforms with values set from the EOSs MM$^-$ (top row) and MM$^+$ (bottom row), and those assuming the qURs. The points of maximum mismatch are shown (black star), with the corresponding masses and mismatch values listed on the upper left. The qURs shown are \textit{Right column}: SIQM, \textit{Center column}: $f$-mode frequency, \textit{Left column}: Binary Love.}
    \label{fig:UR_MM}
\end{figure*}
Contour plots of these mismatches are shown in Fig.~\ref{fig:UR_MM}, along with the points of highest mismatch which are used to simulate waveforms for the selected cases parameter estimation injections. The largest maximum mismatch arises from the SIQM qUR for MM$^+$, reaching $\sim 10^{-1}$, whereas the maximum mismatches from the $f$-mode and Binary Love qURs are smaller, in the range $\sim 10^{-3}$–$10^{-4}$. Maximum mismatches generally occur at lower component masses, reflecting their stronger tidal effects. For the binary Love relation, mismatches are largest for asymmetric binaries and decrease toward the equal-mass limit, as $\Lambda_1 = \Lambda_2$ by construction.
\section{Computing the Spin-Induced Quadrupole Moment}
\label{Appdx:SIQM}

\begin{figure}[htbp!]
        \centering
        \captionsetup{justification=raggedright,singlelinecheck=false}
        \includegraphics[width=0.95\columnwidth]{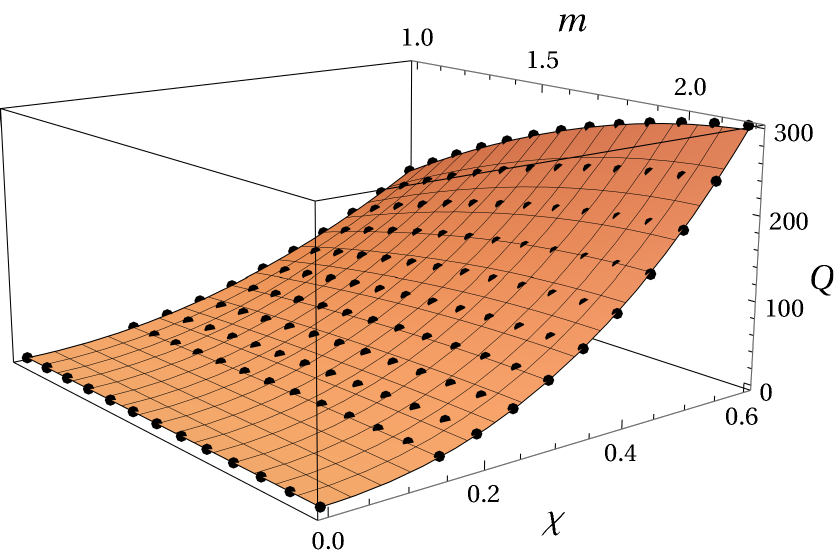}
        \captionsetup{justification=raggedright,singlelinecheck=false}
        \caption{Phenomenological fit for the SIQM $Q=C_Q m^3\chi^2$ (orange surface) alongside the RNS calculated calibration points (black points). }
        \label{fig:Qfit}
\end{figure}

The SIQM was calculated with the \textsc{RNS} code~\cite{Stergioulas:1994ea}, which solves Einstein's field equations for rapidly (uniformly) rotating, relativistic compact stars when supplied with a tabular EOS. This differs from the SIQMs used to calibrate the qUR, which are computed under the slow-rotation approximation~\cite{Hartle:1967he, Hartle:1968si}. We found that RNS failed to converge for low spins. Thus for each EOS, we calculated the SIQM across component masses $m \in [1.0, 2.2]M_\odot$ and spins from the minimum value at which RNS converged for all masses, up to $\chi = 0.6$. These values were then used to construct a phenomenological fit using \textsc{Mathematica}~\cite{Mathematica} with \texttt{NonLinearModelFit} spanning the full spin range, with the boundary condition $Q=0$ at zero spin. Figure \ref{fig:Qfit} shows the phenomenological SIQM fit for MM$^-$ as an illustrative example. The corresponding fit for MM$^-$ is
\begin{equation}
    Q = 507.293m\chi^2 - 44.8039m^3\chi^2-378.495\chi^3 + 315.377m\chi^3
\end{equation}
and for MM$^+$ is 
\begin{equation}
    Q = 568.011m\chi^2 - 335.63\chi^3+62.3359m^2\chi^3 \, .
\end{equation}
SIQM values for all binaries in this study are obtained from these fits.

\section{Calculation of the $f$-mode Frequencies}
\label{Appdx:fmode}

The \textit{f}-mode frequency is calculated following the Lindblom-Detweiler method \cite{1983ApJS...53...73L,1985ApJ...292...12D}, which consists of considering the non-radial pulsations of a compact star in a fully general relativistic context. In this analysis, the governing equations for non-radial oscillations are reformulated as a system of four first-order differential equations
\begin{equation}
\frac{d\mathbf{Y}(r)}{dr}=
\mathbf{Q}(r,\ell,\omega)\mathbf{Y}(r),
\end{equation}
where the quantities \textbf{Q} represent a matrix whose elements depend on $\ell$, and $\omega$, and the equilibrium quantities that can be obtained through the solution of the Tolman-Oppenheimer-Volkoff equations corresponding to the equilibrium configuration. The functions $\mathbf{Y}(r)=(H_1^{\ell m},K^{\ell m},{W}^{\ell m},X^{\ell m})$, correspond to the perturbation variables for the fluid and metric \cite{Flores_2019}. 

We consider the normal modes
with complex frequency
\begin{equation}
 \omega=\sigma+\frac{i}{\tau}.
\end{equation}
where $\sigma$ corresponds to the real part of $\omega$, corresponding to the oscillatory frequency and $\tau$ is the inverse of the imaginary part of $\omega$, corresponding to the damping time os the oscillation.

The normal modes of the coupled system are identified as oscillations that asymptotically generate only outgoing GW at spatial infinity. The real components of their eigenfrequencies specify the oscillation rates, whereas the imaginary components quantify the damping arising from radiative energy losses.

\bibliography{references.bib}

\end{document}